\documentclass[reprint,aps,prapplied,floatfix]{revtex4-2}
\usepackage{graphicx}% Include figure files
\usepackage{dcolumn}% Align table columns on decimal point
\usepackage{bm}% bold math
\usepackage[mathlines]{lineno}% Enable numbering of text and display math
\usepackage[dvipsnames]{xcolor}
\usepackage{amsmath}% For the equation* environment
\usepackage{gensymb}
\usepackage[labelfont=bf]{subcaption}
\usepackage{caption}
\usepackage{xr}
\usepackage{setspace}
\usepackage{hyperref}
\usepackage{amsmath}
\usepackage{physics}
\usepackage{caption}
\usepackage[utf8]{inputenc}
\hypersetup{
  colorlinks   = true, %Colours links instead of ugly boxes
  urlcolor     = blue, %Colour for external hyperlinks
  linkcolor    = blue, %Colour of internal links
  citecolor   = red %Colour of citations
}
\newcommand*{\addFileDependency}[1]{% argument=file name and extension
\typeout{(#1)}% latexmk will find this if $recorder=0
\@addtofilelist{#1}
\IfFileExists{#1}{}{\typeout{No file #1.}}
}\makeatother

\graphicspath{ {./figures/} }
\begin{document}

\preprint{APS/123-QED}

\title{Fast and Sensitive Readout of a Semiconductor Quantum Dot Using an In-Situ Microwave Resonator with Enhanced Gate Lever Arm}

\author{Tim J. Wilson}
 \affiliation{Physics and Astronomy Department, University of California, Los Angeles}
\author{Hong-Wen Jiang}
\affiliation{Physics and Astronomy Department, University of California, Los Angeles}

\date{\today}

\begin{abstract}
We report an experimental study of a Si/SiGe double quantum dot (DQD) directly coupled to a niobium superconducting coplanar stripline (CPS) microwave resonator. This hybrid architecture enables high-bandwidth dispersive readout suitable for real-time feedback and error-correction protocols. Fast and sensitive readout is achieved primarily by optimizing the DQD gate lever arm, guided by MaSQE quantum dot simulations, which enhances the dispersive signal without requiring high-impedance resonators. We demonstrate a signal-to-noise ratio (SNR) of unity with an integration time of $34.54$ ns, corresponding to a system bandwidth of $14.48$ MHz and a charge sensitivity of  $1.86 \times 10^{-4}\, e/\sqrt{\text{Hz}}$ . Analysis of the voltage power spectral density (PSD) of the in-phase (I) and quadrature (Q) baseband signals characterizes the system’s readout noise, with the PSD’s dependence on integration time providing insight into distinct physical regimes.

\end{abstract}

%\keywords{Suggested keywords}%Use showkeys class option if keyword
                              %display desired
\maketitle
\section{\label{sec:introduction} Introduction}

Traditionally, readout in semiconductor quantum dot (QD) devices has been performed using proximal charge sensors, most notably quantum point contacts (QPCs) \cite{wh88,va88} and single-electron transistors (SETs) \cite{av86,fu87}. These devices are typically fabricated in very close proximity to the quantum dot of interest \cite{ko97,ko01,va03,ha07,zw13,bu23}, allowing them to act as sensitive electrometers that detect small changes in the charge configuration of the dot. Over the last three decades, such charge sensors have enabled many of the seminal demonstrations in the field of quantum dots, including the observation of single-electron tunneling, the initialization and manipulation of spin states, and the development of spin qubits. Their sensitivity and robustness made them indispensable tools in establishing the viability of quantum dots as qubit platforms.
However, the requirements of quantum information processing, and particularly quantum error correction, impose more stringent demands on readout. In a scalable quantum processor, qubit states must be measured both rapidly and with high fidelity in order to detect and correct errors before they propagate. The relevant timescales are set by the qubit coherence properties: relaxation processes, characterized by $T_1$ times, and dephasing processes, characterized by $T_2$ and $T_2^*$ times. If measurement requires a time comparable to or longer than these coherence times, the measured state may already have degraded, reducing the fidelity of both state readout and subsequent feedback operations. Unfortunately, conventional QPC and SET sensors are intrinsically bandwidth-limited, with measurement rates typically restricted to tens of kilohertz \cite{vi23}. While this is sufficient for many physics experiments, it is too slow for the demands of large-scale fault-tolerant quantum computing. We will note that Loss-DiVincenzo qubits in purified silicon present extremely long coherence times on the order of a second \cite{st22}, yet with the need for multiple measurements, \cite{ta22}, and the likelihood for feedback protocols. The ability to achieve faster readout still presents a paramount aspect to study, and for systems which are strongly coupled and hybridized systems. Thus, new approaches to readout are required if semiconductor quantum dots are to form the basis of a scalable quantum information platform.
Radio-frequency (rf) reflectometry has emerged as a powerful alternative for charge and spin sensing in quantum dot systems. In this approach, the impedance of the quantum device is probed using microwave-frequency signals, allowing the extraction of state-dependent properties with bandwidths several orders of magnitude higher than conventional dc methods. By embedding the device into a resonant matching network, the reflection coefficient
\begin{equation}
\label{eq:S11}
S_{11}\equiv\Gamma = \frac{Z_\text{load}-Z_0}{Z_\text{load}+Z_0},
\end{equation}
\noindent becomes sensitive to changes in the device impedance, where $Z_\text{load}$ is the load impedance of the device and $Z_0 = 50 \ \Omega$ is the characteristic impedance of the microwave electronics. A practical challenge arises because quantum dot devices typically exhibit resistances on the order of the resistance quantum, $R_Q = h/e^2 = 25.8 \ \text{k}\Omega$. In the limit $Z_\text{load} \gg Z_0$, $\Gamma \to 1$, and the reflected signal carries little information about the device state. Overcoming this impedance mismatch is therefore central to implementing rf reflectometry in quantum devices.
A widely adopted solution has been the use of an impedance-matching tank circuit, typically formed by wirebonding an inductor to the quantum device source contact and exploiting the parasitic capacitance to ground \cite{sc98,re07}. The resulting $LC$ resonator translates the high device resistance into an effective impedance that can be coupled efficiently to a 50-$\Omega$ transmission line. This technique led to the development of rf-QPC and rf-SET devices, which enabled real-time monitoring of charge tunneling and spin dynamics with bandwidths approaching 10 MHz. Both dissipative and dispersive measurements have been demonstrated in such architectures, providing a crucial step toward fast, high-fidelity readout of semiconductor qubits \cite{fr12DI,fr12QD}. Such rf-QPC and rf-SET readout mechanism provide sufficiently fast readout for scalable fault-tolerant quantum computation built upon gated semiconductor quantum dots, yet we would like to demonstrate a system which provides even larger bandwidths.

Building on these advances, a natural next step is to integrate high-quality factor microwave resonators directly on chip and couple them capacitively to quantum dot gates \cite{vi23}. This hybrid approach offers two important advantages. First, it removes the need for dedicated charge sensors and their associated gate structures, simplifying device architectures and improving scalability. Second, it allows direct dispersive readout of the quantum dot states, exploiting their quantum capacitance to shift the resonator frequency in a state-dependent manner. This is analogous to dispersive readout in circuit quantum electrodynamics (cQED), where superconducting qubits are measured via their interaction with microwave resonators.
The coupling strength in such hybrid dot–resonator systems depends sensitively on the lever arm of the gate to which the resonator is coupled. The gate lever arm, defined as
\begin{equation}
\label{eq:LA}
\alpha_\text{g} =\frac{1}{e}\frac{\partial U_\text{well}}{\partial V_\text{g}},
\end{equation}
\noindent quantifies the efficiency with which a change in gate voltage $V_g$ modifies the potential $U_\text{well}$ of the quantum dot. When the resonator electrode and the plunger gate are one and the same, the effective coupling parameter simplifies to $\beta = \alpha_g$, rather than $\beta = |\alpha_r - \alpha_g|$ in split-gate configurations \cite{bo20}. This direct coupling geometry provides the strongest possible interaction between the dot and the resonator field. Compared to impedance engineering, where the coupling strength scales with the square root of resonator impedance, lever-arm engineering offers a linear pathway to stronger coupling.
Optimizing lever arms is therefore a central design goal. Since the lever arm is primarily a geometrical property, it can be predicted and refined using electrostatic simulations of the device structure. In this work, we employ coupled Schrödinger–Poisson simulations to capture both the electrostatics and confinement potential of our Si/SiGe double quantum dot architecture \cite{ca18,gy21}. These simulations allow us to estimate lever arms for different gate geometries prior to fabrication, ensuring that the resonator–dot coupling is maximized while retaining control of the device in the few-electron regime. By aligning design choices with simulation results, we establish a systematic approach to engineering dispersive readout performance.
Once coupled, the double quantum dot–resonator system is naturally described by the Jaynes–Cummings Hamiltonian \cite{jc63},
\begin{equation}
\label{eq:JCH}
H=\hbar\omega_r\left(a^\dagger a+\frac{1}{2}\right)+\frac{1}{2}\hbar\omega_d\sigma_z+\hbar g_\text{eff}(a^\dagger\sigma_-+\sigma_+a ),
\end{equation}
\noindent where $\omega_r$ is the resonator frequency, $\omega_d$ is the qubit frequency of the double quantum dot, $a$ and $a^\dagger$ are photon annihilation and creation operators, and $\sigma$ are Pauli operators describing the DQD pseudospin, and $g_\text{eff}=g_0\frac{2t_c}{\Delta E}$, where $g_0$ is the vacuum Rabi splitting, $t_c$ the tunnel coupling of the DQD, and $\Delta E$ energy splitting between the two-level quantum system. In the dispersive regime, $\omega_d \gg \omega_r$, the dominant interaction manifests as a state-dependent shift of the resonator frequency. This effect can be understood in terms of the quantum capacitance of the DQD,
\begin{equation}
C_Q=(e\alpha)^2\frac{\partial^2 E}{\partial \varepsilon^2},
\end{equation}
\noindent where $E=\sqrt{\varepsilon^2+4t_c^2}$ is the eigenenergy of the DQD within the charge basis, with detuning $\varepsilon$ and tunnel coupling $t_c$ \cite{pe10,vi23}. Changes in $C_Q$ modulate the effective load impedance of the resonator, producing measurable shifts in the reflected signal. This dispersive signature forms the basis of high-bandwidth charge and spin readout, including the detection of Pauli spin blockade \cite{zh19}.

It is possible to connect this microscopic description to experimental observables via input–output theory \cite{co84,bl04,de15,be17,bu20,bl21}. The reflection coefficient of the resonator–DQD system can be expressed as
\begin{equation}
\label{eq:ReflecAmp}
\Gamma\equiv\frac{a_\text{out}}{a_\text{in}}=-\frac{i\delta\omega+\frac{1}{2}(\kappa_\text{int}-\kappa_\text{ext})+g_\text{eff}\chi}{i\delta\omega+\frac{1}{2}(\kappa_\text{int}+\kappa_\text{ext})+g_\text{eff}\chi},
\end{equation}
\noindent where $\delta\omega=2\pi(\nu-\nu_0)$ is the detuning from resonance, $\kappa_\text{int}$ and $\kappa_\text{ext}$ are the internal and external loss rates, and $\chi$ is the susceptibility of the DQD \cite{de15}. The total linewidth $\kappa=\kappa_\text{int}+\kappa_\text{ext}$ determines the loaded quality factor $Q_l=\omega_0/\kappa$ and thus sets the fundamental limit on measurement bandwidth and sensitivity. For this study though, we will not use this method of analysis, and will investigate such an input-output theory model in a future study where we focus on the effects of tunneling events through the DQD on the resonator. In this study, we are primarily focused on the resonator's ability as a readout scheme, and not the connection of the strength between the resonator and the DQD system.

In dispersive readout, the concept of system bandwidth can refer to two physically distinct quantities. The first is the resonator linewidth, $\kappa/2\pi$, which characterizes the rate at which the cavity field relaxes to equilibrium following a perturbation. This quantity sets the intrinsic response time of the resonator–DQD hybrid system and therefore represents a fundamental limit on the temporal resolution of cavity-mediated measurements. In our device, we measure a linewidth of $\kappa/2\pi \approx 2{-}3$ MHz.
The second relevant bandwidth arises from the detection chain and is determined by the finite integration time $\tau$ employed in processing the down-converted $I/Q$ signals. Assuming stationary Gaussian noise, the effective detection bandwidth is $B_\text{eff}=(2\tau_\text{min})^{-1}$.  From our SNR analysis, we extract $B_\text{eff} \approx 14.48$ MHz for an integration time of $\tau = 34.54$ ns. This value does not correspond to the physical linewidth of the cavity but rather to the measurement rate achievable given the amplification chain noise and chosen acquisition parameters.

The charge sensitivity we report, $\sim 2 \times 10^{-4} \frac{e}{\sqrt{\mathrm{Hz}}}$, is thus associated with the detection-limited bandwidth. In other words, the sensitivity quantifies the smallest detectable charge displacement per $\sqrt{\text{Hz}}$ of detection bandwidth, independent of the slower $\kappa/2\pi$ limit imposed by the resonator. The intrinsic linewidth constrains how rapidly the cavity response can follow the DQD state, while the effective bandwidth extracted from SNR analysis reflects how quickly charge state changes can be discriminated with high fidelity. Both must be considered when evaluating the performance of dispersive readout for feedback protocols and error-correction applications.

In this work, we experimentally characterize such a hybrid resonator–DQD system in a Si/SiGe heterostructure. Guided by electrostatic simulations of the device geometry, we optimize the gate lever arm to enhance the dispersive signal. We then demonstrate high-bandwidth readout by achieving a signal-to-noise ratio of unity at nanosecond integration times, corresponding to charge sensitivities well suited for real-time feedback and future error correction protocols. By analyzing noise spectra extracted from the baseband in-phase and quadrature signals, we further characterize the charge noise environment of the device, linking observed $1/f$ noise behavior to charge fluctuations within the double quantum dot. Together, these results establish resonator-coupled quantum dots with engineered lever arms as a scalable pathway toward fast, high-fidelity qubit readout.

In this study, rather than modifying the resonator impedance, we focus on maximizing the lever arm, $\alpha_g$, through device design. To this end, we employ self-consistent electrostatic simulations of the coupled Schrödinger–Poisson equations in the Si/SiGe heterostructure to extract an estimated lever arm for a given gate geometry. These simulations allow us to optimize gate placement and confinement potential prior to fabrication, thereby ensuring strong capacitive coupling between the DQD charge states and the cavity electric field. Our simulation results indicated this device design would have about a two times larger lever arm than our typical depletion mode devices, from $\sim 0.1$ to $\sim 0.2$ \cite{ro25}.

A stronger couple between a QD system and a resonant circuit has been primarily facilitated in the community by increasing the resonant circuits impedance, $Z_r$. By employing either a geometrically defined high impedance coplanar waveguide (CPW) \cite{mi17,mi18,sa18,un24}, or a high kinetic inductance system such as  superconducting quantum interference device (SQUID) arrays \cite{sa16,sc22}. These studies are primarily focusing on achieving the strong coupling regime, which requires an optimization of the vacuum Rabi splitting of the system which is directly related to the resonant circuit's impedance. This is as the coupling strength; the vacuum Rabi splitting, is related by: 

\begin{equation}
    \frac{g_0}{2\pi}=\frac{\alpha}{2}\nu_0\sqrt{\frac{Z_0}{\pi \hbar}}
    \label{eq:VRS}
\end{equation}

\noindent thus, a direct increase in the vacuum Rabi splitting is most easily attained with an increased resonator impedance. There is also a dependence on the gate lever arm, $\alpha$, and we will discuss our reasoning to choose this parameter to maximize.

This study is focused on the dispersive regime, where the DQD–cavity detuning satisfies $\Delta = |\omega_a - \omega_0| \gg g_\text{eff}$. In this limit, the Jaynes–Cummings Hamiltonian can be expanded in powers of the small parameter $g/\Delta$ \cite{bl04}, yielding an effective dispersive Hamiltonian of the form

\begin{equation}
H = \hbar\left(\omega_0 + \frac{g^2}{\Delta}\sigma_z\right)\left(a^\dagger a + \frac{1}{2}\right) + \hbar\omega_a\frac{\sigma_z}{2}.
\label{eq:DRJCH}
\end{equation}

The central feature of Eq.~\eqref{eq:DRJCH} is that the cavity frequency acquires a state-dependent shift $\chi = g^2/\Delta$, which effectively imprints the DQD charge state onto the microwave field. Measurement of the transmitted or reflected microwave field then provides information about the DQD without the need for direct current transport. This dispersive frequency shift is accompanied by a modification of the cavity linewidth through the Purcell effect, leading to both frequency pulling and linewidth broadening that depend on the detuning and the charge hybridization of the DQD.
Dispersive readout thus constitutes a quantum non-demolition (QND) measurement scheme: the DQD state modulates the resonator properties without inducing direct transitions between eigenstates, provided the measurement drive remains sufficiently weak; $\frac{g}{\Delta}\ll 1$. This makes the technique particularly appealing for spin–photon and charge–photon coupling experiments, where preserving coherence while extracting state information is critical \cite{mi17}. In the context of semiconductor DQDs, dispersive readout offers several specific advantages. First, it avoids the need for dc transport, enabling state readout even in regimes where source–drain coupling is suppressed or undesirable. Second, and most importantly for our study, it allows for high-bandwidth measurements limited primarily by the resonator linewidth \cite{st15}, which in our devices is on the order of 2–3 MHz. Finally, the non-demolition character of the measurement makes dispersive readout naturally compatible with future qubit encoding schemes where repeated measurements and error correction will be essential \cite{ba14}.

Within the dispersive regime of cQED, in the case of a resonant cavity coupled to a DQD, the goal would be to maximize the sensitivity of the cavities' response to changes in the DQD state. This sensitivity would typically be characterized by the charge susceptibility of the DQD system, $\chi=\frac{\partial <n>}{\partial \varepsilon}$. The gate lever arm, Equation \ref{eq:LA}, quantifies how strongly changes of the associated gate voltage affect the energy levels in the QD system. Thus, a larger lever arm directly enhances the charge susceptibility, which allows for a greater control of modulating the resonators characteristics based on the DQD state. That is by being able to have a more significant effect on the qubit frequency, $\omega_q$, one can have a sharper response in the reflectometry measurement. If one considers, $\frac{\partial \chi}{\partial V_g}=\frac{\partial \chi}{\partial\varepsilon}\frac{\partial\varepsilon}{\partial V_g}=e\alpha_g\frac{\partial \chi}{\partial\varepsilon}$, then this effect becomes clear. The lever arm directly relates to the variation of the charge susceptibility with respect to the gate voltage applied to the device. Thus,  here we focus on enhancing the lever arm, $\alpha$, leading to an enhancement of the charge susceptibility, $\chi$, which in turn gives us an increase in the sensitivity of the resonator's response to the DQD state \cite{vi23}.

Enhancing the resonator impedance, $Z_r$, leads to an enhancement of the electric dipole coupling directly influencing the effective coupling strength, $g_\text{eff}$, in the Jaynes-Cummings Hamiltonian, Equation \ref{eq:JCH}. In our study though, maximizing $\alpha_g$ not only increases the resonator response to charge fluctuations but also enhances sensitivity to low-frequency noise sources intrinsic to the DQD as well. Thus, by optimizing the lever arm, the same measurement chain that provides high-bandwidth dispersive readout also enables quantitative characterization of charge noise through voltage PSD analysis of the $I/Q$ signals.

Dispersive readout represents a quantum non-demolition measurement scheme and is thus very appealing for scalable architectures \cite{bl04}. Within this regime, the goal is to maximize the sensitivity of the cavity response to changes in the DQD state. This sensitivity is characterized by the charge susceptibility of the DQD system, $\chi = \frac{\partial \langle n \rangle}{\partial \varepsilon}$.
The gate lever arm, Equation \ref{eq:LA}, quantifies how strongly changes in the applied gate voltage affect the QD energy levels. A larger lever arm directly enhances the charge susceptibility, thereby increasing the ability of the DQD to modulate the resonator properties. This effect can be expressed as
\begin{equation}
\frac{\partial \chi}{\partial V_g}
= \frac{\partial \chi}{\partial \varepsilon} \frac{\partial \varepsilon}{\partial V_g}
= e \alpha_g \frac{\partial \chi}{\partial \varepsilon}.
\end{equation}
Thus, a larger lever arm, $\alpha_g$, allows for sharper reflectometry signatures through stronger modulation of the effective qubit frequency, $\omega_q$.
In contrast to approaches based on enhancing the resonator impedance, $Z_r$, here we optimize the lever arm, $\alpha_g$, through device design. Specifically, we employ self-consistent electrostatic simulations of the coupled Schrödinger–Poisson equations in the Si/SiGe heterostructure to extract the lever arm for candidate gate geometries. These simulations allow us to preselect designs with maximized capacitive coupling between the DQD charge states and the cavity electric field, ensuring strong dispersive signatures in experiment. It further allows for the important aspect of the potential landscape to provide a quantum dot system to form. That is, we can visualize the charge confinement via the result of the charge density of the simulation's result.

After we can get a good confirmation of charge confinement with our device geometry, we are interested in finding the lever arm of the gates. The lever arm of the gates tells one how the potential landscape of the quantum well changes in response to changes in the voltage of the gates, thus ultimately affecting the chemical potential in the system. The lever arm is a conversion factor from the well's potential to the gate's voltage \cite{va03}. This is defined as: $\alpha_{G_i}\equiv \frac{1}{e}\frac{\partial U_{\mathrm{well}}}{\partial V_{G_i}}$, where $e$ is the elementary charge, $U_{\mathrm{well}}$ is the electrochemical potential of the well, $V_{G_i}$ is the voltage associated with the gate, $G_i$, where $i$ indexes the total number of gates. The lever arm is important for this particular type of device design which is coupling a resonator to the DQD. That is because the coupling strength, $g$, between the photon and the energy levels of the QD is directly proportional to the lever arm, Equation \ref{eq:VRS}; $g \propto \alpha$ \cite{vi23, bo20}. Furthermore, the signal-to-noise ratio of such a system scales as the cube of the lever arm; $\mathrm{SNR}\propto \alpha^3$ \cite{bo20}. From these considerations it is quite clear we would want to maximize the lever arm as much as possible for our design.

If we consider the definition of the lever arm, then we see that it implies for the physical system that the area of the gate should be large. That is since the size of the gate directly relates to its affect of the potential landscape. It has been seen that the size of the dot directly relates to the number of charge carriers it traps and further affects energy level spacings \cite{ci00, va03}. But we can only make these gates so large before not being able to confine only a few electrons. The effective mass also plays a role in this, as smaller effective masses allow for less stringent nanofabrication \cite{bu23}. For certain systems this is easier, say, in GaAs systems, since the effective mass for electrons is so low, one can form very large gates to confine a few electrons without much issues as compared to Si systems. Within our group, GaAs QDs were formed with length scales of a few-several $ 100$ nm \cite{zh09,ho09,ho13}, whereas for Si systems the length scales have been $\sim 100 $ nm \cite{xi10,pa12,ha14,fr16,sc17,pe19}. So one needs to consider the effective mass of the charge carriers in order to determine dot sizes. It has also recently been seen that SiGe/Ge, where the QW is Ge, allows for larger dots as well due to the very low effective mass of the charge carriers exploited in such heterostructures \cite{lo19,he20}. A recent study in our group have been fabricated in Ge QW heterostructures, where the length scale was $\sim 150$ nm \cite{ro25}. But in an Si system we have a much larger effective mass for the electron and the confinement area must be smaller. Here we will discuss our efforts to maximize the lever arm for our design while maintaining good confinement and a potential landscape which allows good manipulation of the system for the experiments we desire to carry out.

In order to find the lever arm for one gate, we would need to first have a reference simulation. That is, we choose potentials of the gates which leads to a seemingly good state for well defined quantum dots. For a DQD we take this as a configuration where there is $\sim 1$ electron in each dot, this is our reference point. Now we select one of the gates and vary its potential by a small amount,  $5$ meV. We need this to be a small quantity in comparison to the total potential of the gate as we are approximating a differential with a finite difference. This gives us a second simulation with an altered potential landscape due to the change in potential of one of the gates; a comparison simulation. The difference in the potential of the reference simulation to that of the comparison simulation gives us an approximation of the change in the potential of the well, that is $\partial U_{\mathrm{well}}$. This quantity divided by the energy change in the gate, that is $\partial V_{G_i}$, gives us the lever arm associated with that gate. Since both these quantities are given to us in units of eV, we can ignore the factor of elementary charge, $e$, in the definition of the lever arm. Since there will be multiple gates for any such system, there will be a lever arm for each gate. Thus, for a system with $M$ gates, we must have $M+1$ total complete simulations to attain the $M$ lever arms.

It is important to distinguish between two relevant bandwidths in this context. The resonator linewidth $\kappa/2\pi \sim 2{-}3$ MHz sets the intrinsic response rate of the cavity–DQD system. However, the effective detection bandwidth, extracted from SNR analysis, reflects the rate at which state discrimination can be performed for a given integration time $\tau$. In our measurements, we observe an effective bandwidth of $14.48$ MHz for $\tau = 34.54$ ns, which exceeds the bare linewidth.
Finally, we note that maximizing $\alpha_g$ also increases sensitivity to charge fluctuations, enabling characterization of low-frequency noise. The same high-bandwidth dispersive measurement chain used for state readout can therefore be applied to extract the charge noise spectrum via power spectral density analysis of the $I$ and $Q$ quadratures, providing insight into the microscopic noise processes in Si/SiGe DQDs.

In our system, the effective coupling between the cavity and the DQD does not arise directly from the bare dipole interaction, but rather through the gate electrode that mediates the resonator field onto the DQD detuning axis. The bare dipole coupling to an electric field mode of the resonator can be written as:

\begin{equation}
    g=\frac{eE_\text{zpf}d}{\hbar}
\end{equation}

\noindent where $E_\text{zpf}$ is the zero-point electric field fluctuations of the resonator and $d$ is the effective dipole length of the DQD charge transition. However, the resonator is capacitively coupled to a gate electrode, so that its voltage fluctuations do not directly act on the orbital wavefunctions but instead shift the detuning $\varepsilon$ through the gate lever arm $\alpha_g$. Specifically, a gate voltage shift $\delta V_g$ modifies the detuning according to:

\begin{equation}
    \delta\varepsilon=\alpha_g\delta V_g
\end{equation}

For the quantum fluctuations of the cavity field, the relevant scale is the resonator zero-point voltage fluctuations $V_\text{rms}$, such that:

\begin{equation}
    \delta\varepsilon=\alpha_g V_\text{RMS}
\end{equation}

This results in an effective DQD–cavity interaction strength

\begin{equation}
    g_\text{eff}=\frac{e\alpha_g V_\text{RMS}}{\hbar}
\end{equation}

Thus the coupling strength is set not only by the bare resonator vacuum fluctuations, but also by the electrostatic conversion factor between gate voltage and dot detuning, i.e. the lever arm. This gate-dependent renormalization explains why the experimentally relevant coupling is best described as

\begin{equation}
    g_\text{eff}=\alpha_gg_0
\end{equation}

where $g_0$ denotes the bare resonator coupling to the gate voltage and $\alpha_g$ quantifies the geometric efficiency of converting this voltage into detuning. Consequently, the dispersive shift $\chi = g_\text{eff}^2 / \Delta$ depends sensitively on the device geometry through $\alpha_g$. This highlights the importance of optimizing the capacitive coupling between the resonator field and the DQD plunger gate: a larger lever arm directly enhances the dispersive shift, thereby improving the measurement signal-to-noise ratio without requiring stronger drive powers that could induce backaction.
Moreover, because the lever arm is determined by the device layout and material stack (oxide thickness, gate pitch, and screening from neighboring electrodes), our fabrication choices directly impact the achievable readout fidelity. For example, the thin ($< 5$ nm) Al$_2$O$_3$ oxide layer deposited in the active DQD region not only minimizes unwanted fringe fields but also enhances $\alpha_g$, allowing the plunger gates to more efficiently modulate the DQD detuning. Thus, the dispersive readout characteristics observed in our experiments can be directly traced to both the effective Hamiltonian of Eq.\eqref{eq:DRJCH} and the specific device engineering steps described in Sec.\ref{ssec:devicefabrication}.

\section{\label{sec:setup} Experimental Setup and Device Under Test}

\begin{figure}
    \centering
    \includegraphics[width=\linewidth]{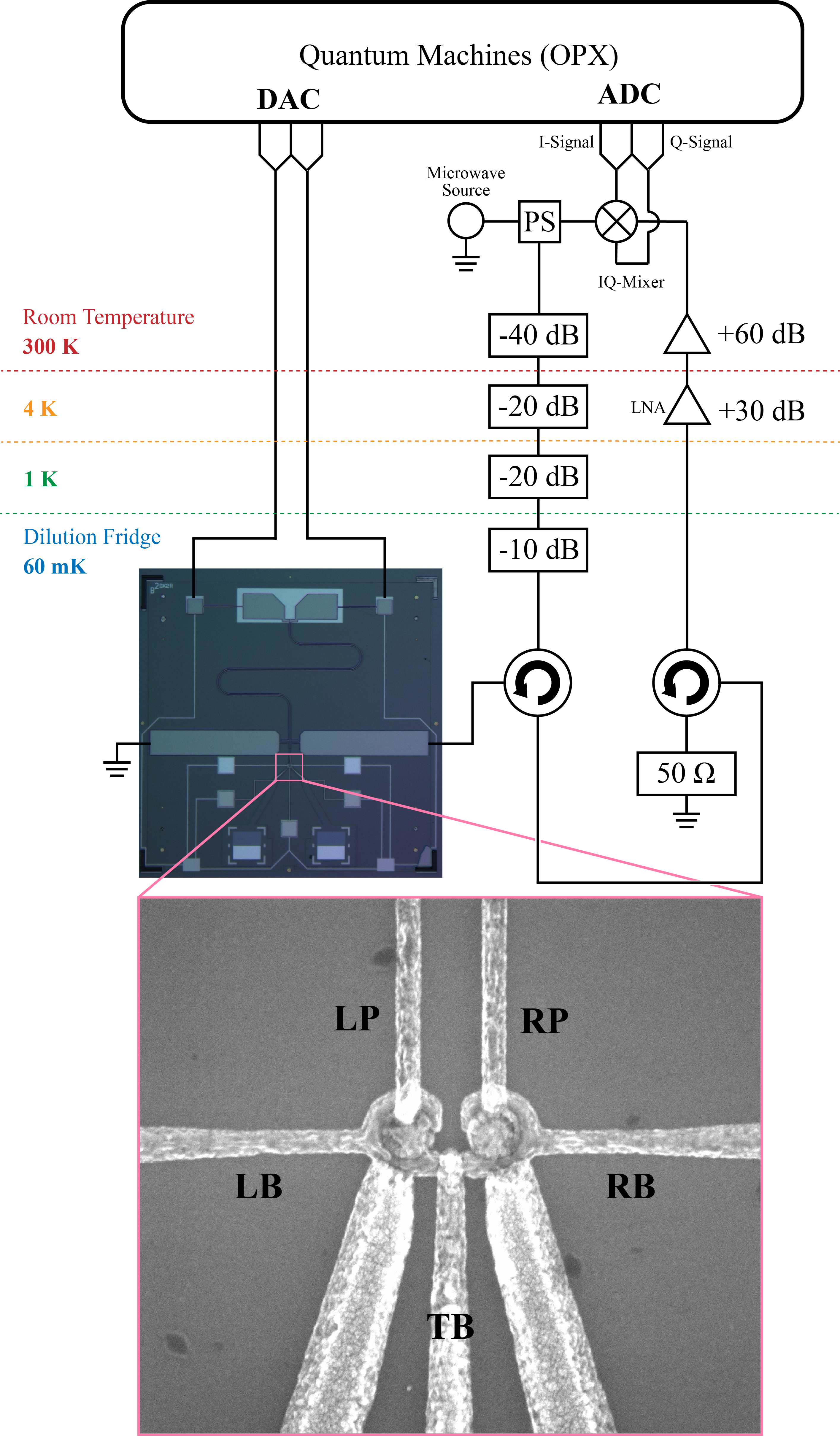}
    \caption{Illustration of experimental setup. Experiments were conducted in an Oxford Instruments Triton dilution fridge with base temperature of 60 mK. The microwave generator signal is split to device and IQ-mixer. Cold stage attenuation is used to avoid thermal noise, which could lead to qubit dephasing and thermal broadening. The single port systems is made into a two port measurement setup by using a circulator to send the input microwave signal to the device and the output reflected signal can be amplified and compared to the input; isolating reflections. We measure the real ( I ) and imaginary ( Q ) part of the reflection coefficient, $S_{11}$ and apply DC offsets to the plunger gates using a quantum machines OPX. A 4K stage low noise amplifier (LNA), by low noise factory is used to amplify the signal from the devices, this is critical to attain a signal. The labels on the device zoom-in correspond to L for left, R for right, P for plunger, T for tunnel, and B for barrier.}
    \label{fig:ExpSU}
\end{figure}

Our experiments are conducted in an Oxford Instruments Triton dilution fridge with a mixing chamber at 60 mK. A combination of rigid and semi-rigid high frequency lines are used to send microwave/rf signals to the device and amplify the reflected signal from the device. We power the device and the IQ mixer with the same microwave signal, thus we must attenuate the signal before it reaches our device under test (DUT), as the IQ mixer requires 10 dBm to operate. We use attenuators at various stages to avoid a large thermal load in our dilution fridge as attenuators are lossy forms of power dissipators. The higher values of attenuation are used at higher temperatures to avoid the thermal noise associated with them. An illustration of our setup can be seen in Figure \ref{fig:ExpSU}.

\begin{figure}
    \centering
    \includegraphics[width=\linewidth]{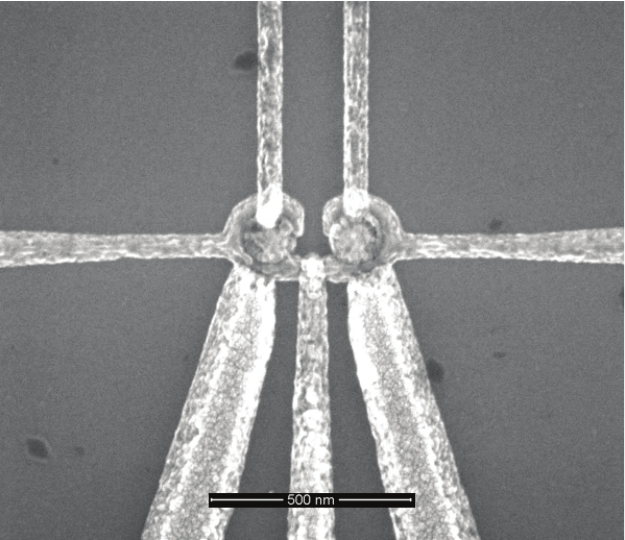}
    \put(-85,138){\includegraphics[width=0.33\linewidth]{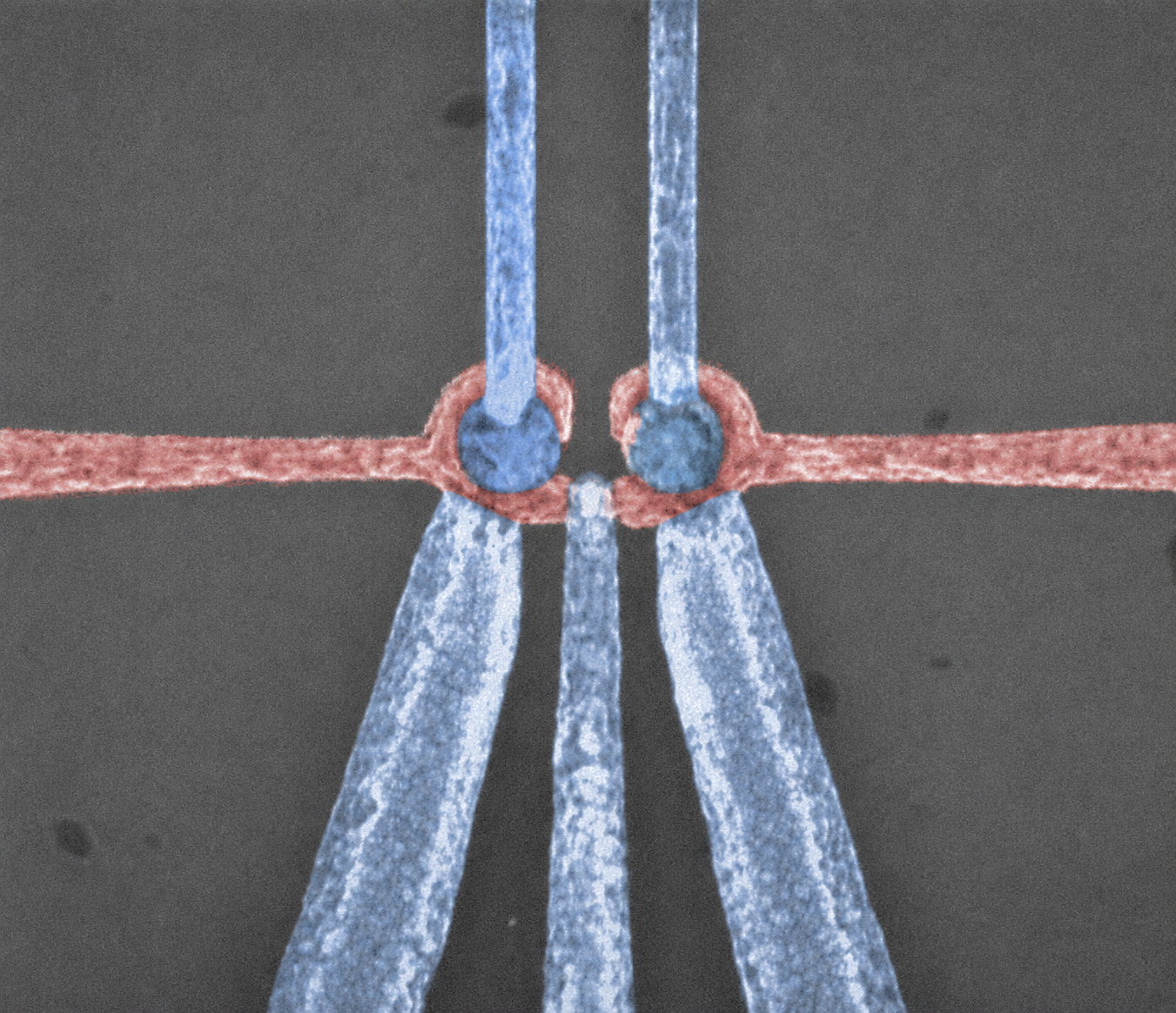}}
    \caption{SEM of device used for the experiments. The design is an accumulation mode device with overlapping gates. Inset figure shows a false color of the device to illustrate the two layer overlapping gate design where the red false color is patterned and deposited first, and the blue false color is patterned and deposited afterwards.}
    \label{fig:DevSEM}
\end{figure}

The device we use was fabricated on a Si/SiGe smaple provided to us by Hughes Research Laboratories (HRL). The heterostructure was grown on a Si substrate of $\sim 500  \ \mu$m, with a Si$_{0.7}$Ge$_{0.3}$ buffer layer of $225$ nm, a strained Si QW of $5$ nm, and a Si$_{0.7}$Ge$_{0.3}$ cap of $50$ nm. The gate geometry is of the overlapping design with two layers, this was done primarily to make a more cost effective fabrication procedure for overlapping gate devices. There is a caveat to doing only two layers as the barrier gates in this design must act not only to provide a barrier height for the potential islands for the QDs, but it must also act as a screening channel. This has proved some challenge in fine tuning the DQD device. A scanning electron microscope (SEM) image of the device can be seen in Figure \ref{fig:DevSEM}

\begin{figure*}
    \centering
    \includegraphics[width=0.95\textwidth]{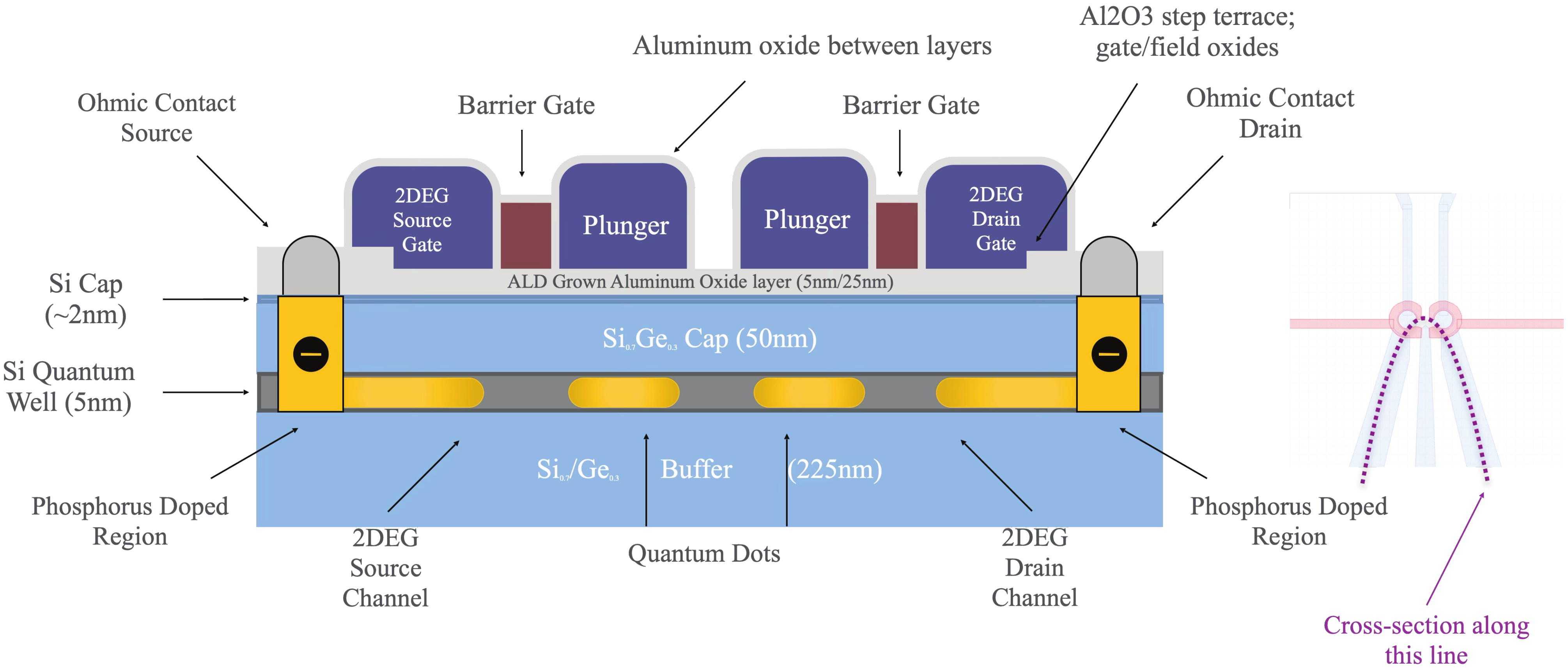}
    \caption{Schematic of the heterostructure and device design.}
    \label{fig:HetSch}
\end{figure*}

\section{\label{sec:results}Results}

We characterize a superconducting coplanar strip-line (CPS) niobium resonator, which is directly connected to the plunger gates of a double quantum dot system in a Si/SiGe heterostructure. The characterization involves the evaluation of the resonator's key parameters, the coupling of the resonator to the double quantum dot system, the charge noise and charge sensitivity associated with using the resonator as a sensor, as well as the signal-to-noise ratio of the system and its minimum integration time. We do this in the dispersive regime of the system, where the two-dimensional electron gas (2DEG) passing through the DQD channel acts as a capacitive loading of the resonator.

\subsection{\label{ssec:device} Device Design}

The resonator and DQD are patterned on a Si/SiGe heterostrcuture. The device is an overlapping gate geometry structure operating in accumulation mode. Two layers define the gates and this was done to have a cost effective design to more easily fabricate than the traditional three layer overlapping gate designs. Furthermore, the gate layout was specified with as few gates as possible for ease of control and to have as few DC lines as possible. The two plungers gates for the DQD connect directly to the two conducting strips of Nb that form the CPS resonator.

\subsection{\label{ssec:devicefabrication} Device Fabrication}

The device was fabricated on a Si/SiGe heterostructure consisting of a $5$ nm strained Si quantum well (QW) grown atop a $225$ nm Si$_{0.7}$Ge$_{0.3}$ buffer and capped by $50$ nm of Si$_{0.7}$Ge$_{0.3}$. To prevent parasitic loading of the coplanar stripline (CPS) resonator by a two-dimensional electron gas (2DEG), we etched down just below the QW in regions where the resonator and bond pads were defined. This etching was performed in a reactive ion etcher.
Ohmic contacts were formed by phosphorus ion implantation into lithographically defined windows, followed by rapid thermal annealing at $750^\circ$C for 30 s to repair lattice damage. After removing the resist, $50$ nm of Al was deposited by e-beam evaporation onto the Ohmic windows, following a 15 s buffered oxide etch to remove the native SiO$_2$. A forming gas anneal at $420^\circ$C for $\sim 30$ min promoted dopant activation and diffusion.
Device borders and the base plate of the CPS termination capacitor were then defined in a second metallization step. The wafer was subsequently coated with a 20 nm layer of Al$_2$O$_3$ grown by atomic layer deposition (ALD). This dielectric was selectively etched in the DQD active region, where a thinner ($< 5$ nm) ALD Al$_2$O$_3$ layer was regrown to suppress fringing fields from the gates.
Fine gate electrodes were patterned in two electron-beam lithography (EBL) steps. The first layer consisted of $50$ nm of Al; after patterning, the sample was exposed to an oxygen plasma for 30 min in a plasma asher to improve adhesion and native oxide quality. The second gate layer was metalized with 65/5 nm of Al/Pt. The Pt capping layer prevents oxidation and ensures electrical continuity with the subsequently deposited CPS resonator.
Finally, the CPS resonator was patterned by photolithography and sputtered with 100 nm of Nb, overlapping the Al/Pt plunger gate leads to complete the device.

\subsection{\label{ssec:Resonator} Resonator}

The resonator is a $\frac{\lambda}{4}$ CPS comprised of two conducting strips. The two lines are each $10 \ \mu$m with a gap separation of $8 \ \mu$m. An node is created by a termination capacitor, $C_T\sim 100$ pF, fabricated on chip, which is a short for microwave frequencies, but open to DC which is where the plungers gates receive their DC bias from. The antinode is located at the center of the plunger gates with a total length of $l_\text{CPS}=4.63$ mm. This design is not common in the community, but we have had collaborations with another group using a very similar design \cite{sc14}. It is patterned with Nb which is superconducting in our experimental setup. From a purely geometric consideration of the resonator, we can assign an effective dielectric constant of $\epsilon_\text{eff}=6.35$, a resonator line impedance of $R_r=89.8 \ \Omega$, an inductance per unit length of $L_l=0.754 \ \mu$H/m, a capacitance per unit length of $C_l=93.7$ pF/m, resulting in a lowest order mode resonant frequency of $\nu_r=\frac{1}{4l_\text{CPS}\sqrt{L_lC_l}}=6.42$ GHz.

We experimentally extract the in-phase, $I$, and quadrature, $Q$, signals from an IQ-mixer. The IQ-mixer is our reflectometry circuit as the in-phase signal corresponds to the real part of the reflection coefficient and the quadrature corresponds to the imaginary part; $I\sim \Re(S_{11})$, $Q\sim \Im(S_{11})$. We are measuring a baseband $IQ$ signal, extracting the components directly from a single IQ-mixer that is down-converting the resonator output to DC up to a few MHz. The signal from the IQ-mixer is feed into the analog inputs of a FPGA controller made by quantum machines to be measured. This instrument allows for fast DC biases to be applied with $16$ ns pulse lengths and similarly can record an ADC signal with a minimal integration time of $16$ ns.

The measured resonant frequency is $\nu_r=5.9665$ GHz and we find a loaded quality factor of $Q_l=3000$ when the QD system is not tuned, and no 2DEG is accumulated near the resonator from the gates. When the system has a 2DEG accumulated by gates near the resonator we find the quality factor to vary from 1000-2000. The likely reason for such a low quality factor with an Nb superconducting resonator seems to stem from an unintentional accumulation of a 2DEG under the portion of the CPS that still has the QW underneath it. Though this loaded quality factor could hinder long distance coupling experiments \cite{pe12, bo20}, this device architecture would still present itself well for long range photon mediated qubit-qubit coupling if strong coupling could be demonstrated. Our focus is in fast read out and this lowered quality factor comes into our benefit as the system bandwidth is inversely proportional to the quality factor of the system.

We characterize the resonators coupling to the 2DEG by looking for a dispersion shift in our resonator signal when there is considerably tunneling events through the DQD channel. These tunneling events through the DQD correspond to a capacitive loading which pulls the resonator frequency shifting it downward.

\begin{figure*}
    \centering
    % Row 1
    \begin{subfigure}{0.5\textwidth}
        \includegraphics{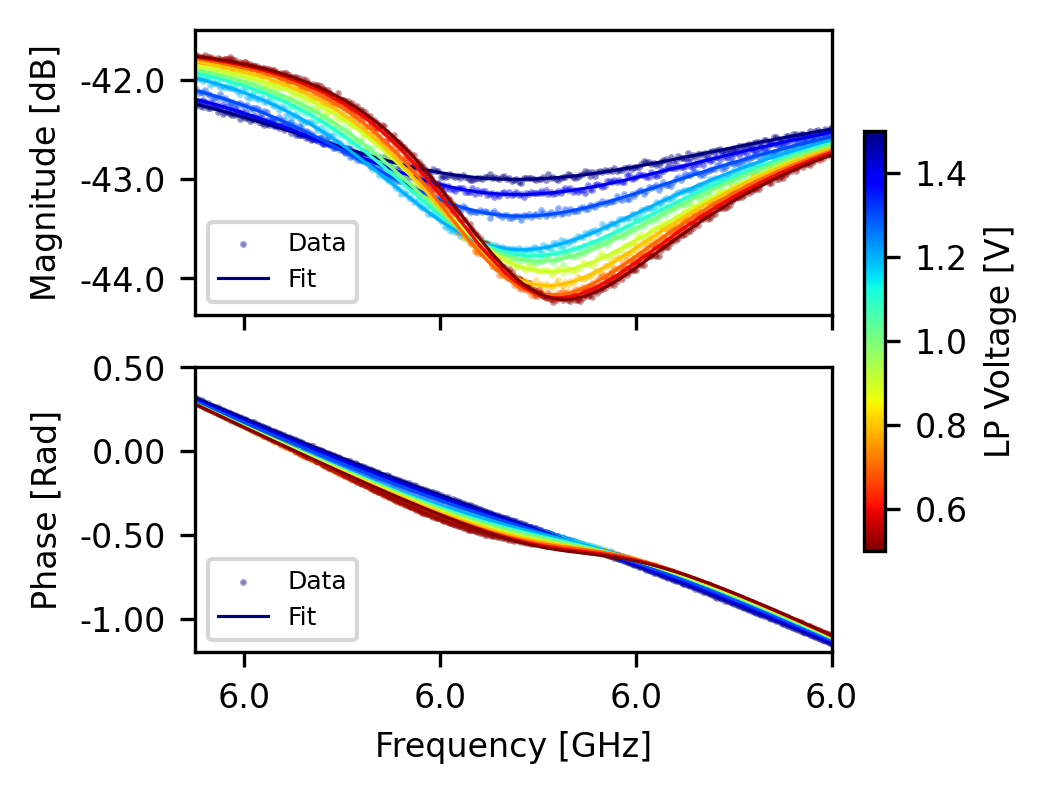}
        \caption{\label{fig:PLMathFits}}
    \end{subfigure}\hfill
    \begin{subfigure}{0.5\textwidth}
        \includegraphics{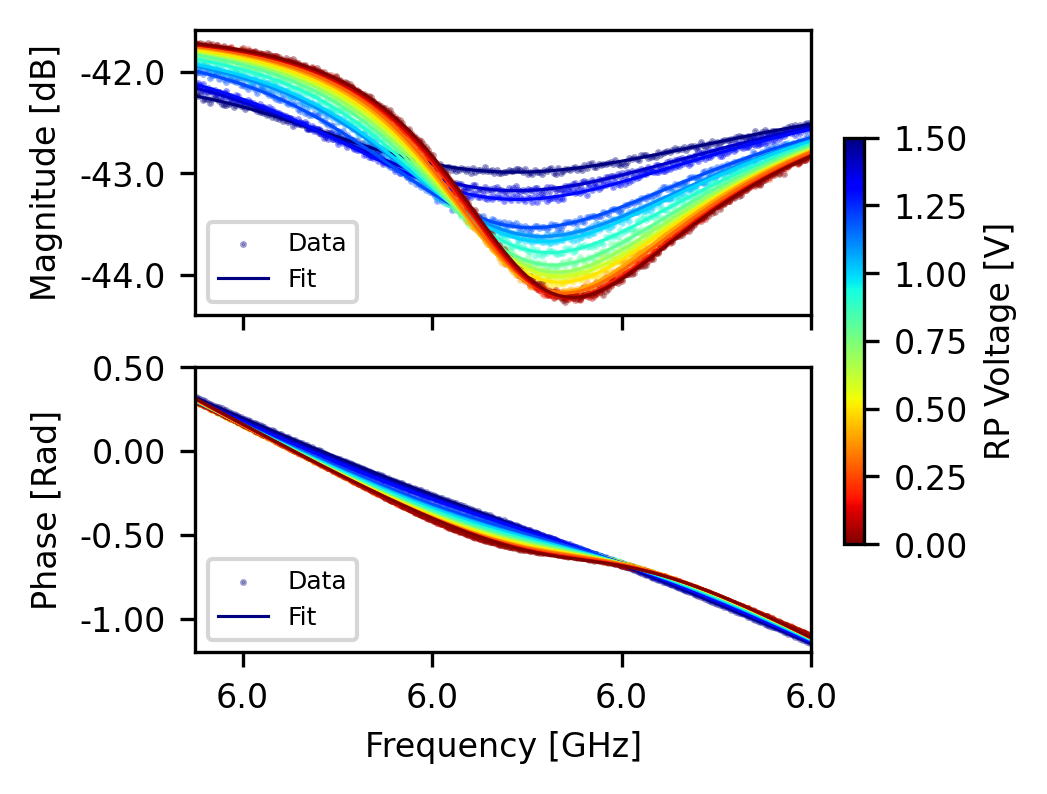}
        \caption{\label{fig:PRMathFits}}
    \end{subfigure}

    % Row 2
    \begin{subfigure}{0.5\textwidth}
    \includegraphics{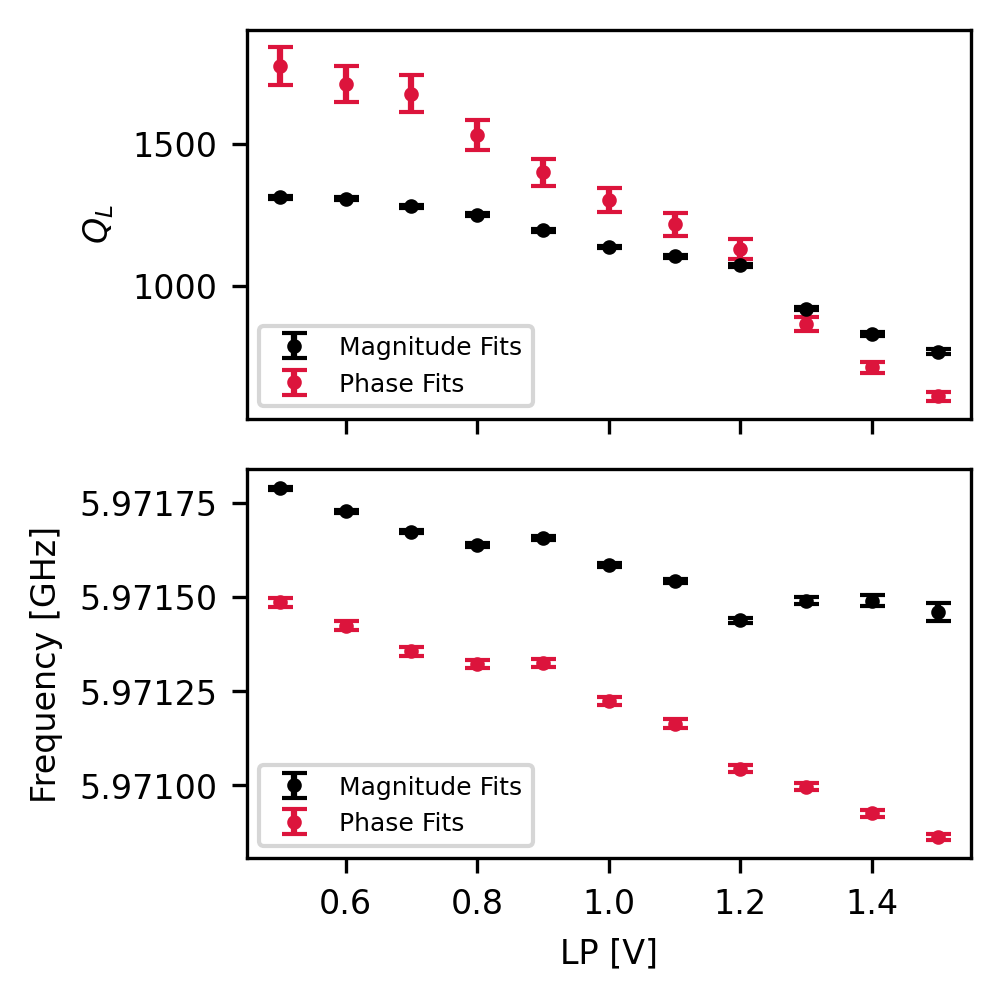}
        \caption{\label{fig:PLfitparams}}
    \end{subfigure}\hfill
    \begin{subfigure}{0.5\textwidth}
    \includegraphics{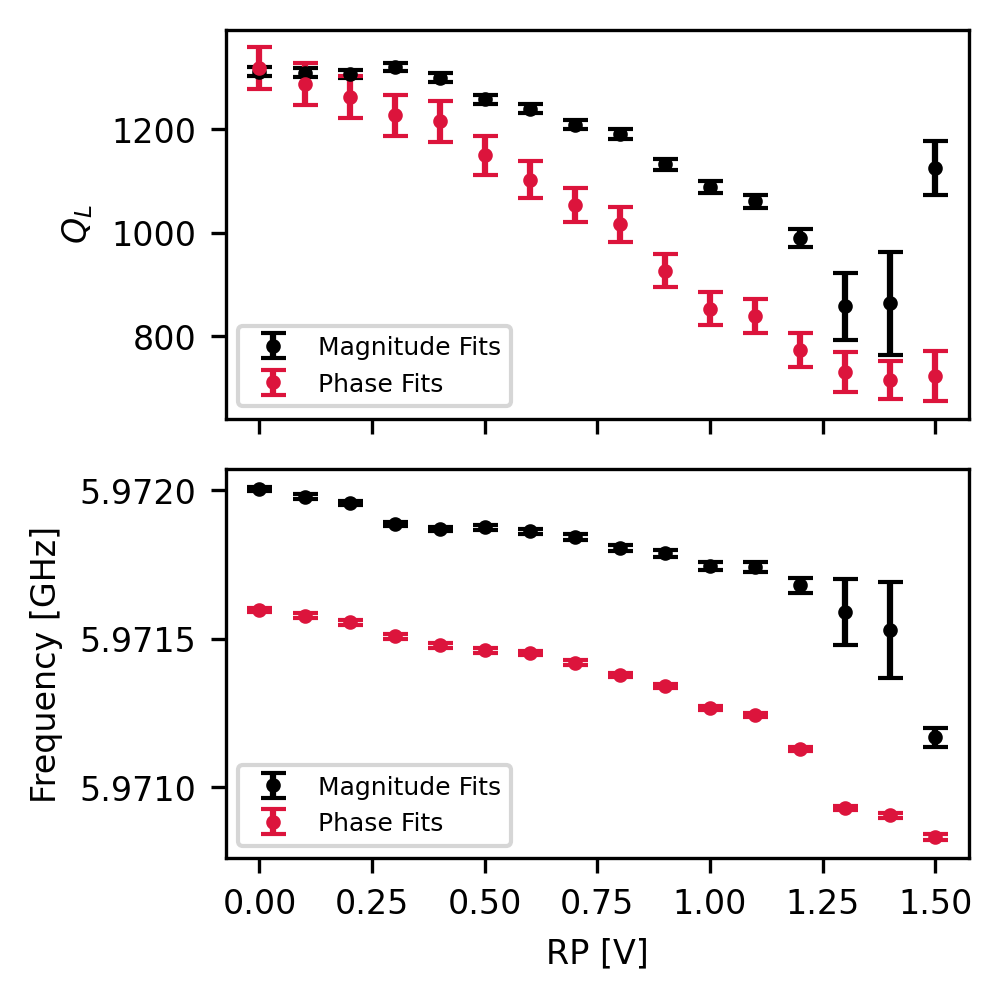}
        \caption{\label{fig:PRfitparams}}
    \end{subfigure}\hfill

    \caption{\label{fig:device} 
      \textbf{(a)} Measured reflection coefficient magnitude and phase as a function of the left plunger gate voltage as well as mathematical fits described by Equations \ref{eq:voigt} and \ref{eq:tanphase}.  
      \textbf{(b)} Same as panel \textbf{(a)} but as a function of the right plunger gate voltage.  
      \textbf{(c)} Loaded quality factor and resonant frequency based on fits for left plunger gate value sweep.
      \textbf{(d)} Loaded quality factor and resonant frequency based on fits for right plunger gate value sweep.}
      \label{fig:FitResults}
\end{figure*}

We observe a dispersive shift of the resonant frequency due to loading of the resonant from the 2DEG density passing through the QD channel. We can force a strong dispersive shift of the resonator by operating the DQD device as a MOSFET channel where there is a steady current flowing through the DQD channel. The amount of current flowing through the channel gives a controllable modulation of the quality factor of the system. This can be seen in Figure \ref{fig:FitResults}, where the resonant frequency is shifted and quality factor modulated as a function of each plunger gate's voltage. We fit this data in two ways, one by a mathematical model and the other by a physical model. The mathematical model is by fitting the resonance curve measured to a pseudo-Voigt function which is a convolution of a Gaussian function, $G(\nu)$ and a Lorentzian function $L(\nu)$:

\begin{equation}
	\label{eq:voigt}
	M(\nu) = (1-f)G(\nu) +fL(\nu)
\end{equation}

\noindent where $f$ is the character fraction detailing a percentage of the curve being Gaussian or Lorentzian, when it is unity the fit is purely Lorentzian, when it is zero the fit is purely Gaussian. We further use a smooth sigmoid logistic function in place of the full width at half maximum, $\gamma$, to account for asymmetry of the curve. This fit, though not derived by the physical nature of the system, gives us physically significant quantities for the resonant frequency, $\nu_0$, and the loaded quality factor, $Q_l=\frac{\nu_0}{\gamma}$. We then separately fit the phase to:

\begin{equation}
	\label{eq:tanphase}
	\varphi(\nu)=A\arctan[2Q_l(\nu-\nu_0)]+B\nu+C
\end{equation}

\noindent where $A$, $B$ and $C$ are arbitrary free parameters in order to fit the measured data and a linear frequency term is incorporated to account for drift in the phase. These two fits independently give us the resonant frequency and quality factor of the loaded system, the agreement of these two fits can be seen in Figure \ref{fig:FitResults}.

For our system we are operating around -100 dBm to -110 dBm, corresponding roughly to $10^2-10^3$ photons in the system given the relation $\langle n \rangle = \frac{P_{\text{in}}}{\hbar \omega_0}  \frac{\kappa_{\text{ext}}}{\left(\delta\omega^2+\frac{\kappa^2}{2}\right)}
$.

\subsection{\label{ssec:LA} Lever Arms}

The lever arm relates the voltage applied to a gate electrode to its effect on the electrochemical potential seen by the QD \cite{va03}. A matrix form of the lever arm follows from the definition:

\begin{equation}
    \vec{\mu} = e \boldsymbol{\alpha} \vec{V}
\end{equation}

\noindent where $\vec{\mu}$ is the column vector of the electrochemical potential of the DQD, $e$ is the elementary charge,  $\boldsymbol{\alpha}$ is the lever arm matrix, and $\vec{V}$ is the vector of gate voltages applied to each gate electrodes considered. The matrix elements of the lever arm are as follows, and gives the dimensions of the matrix: $\alpha_{ij}\equiv \frac{1}{e}\frac{\partial U_{\mathrm{d}_i}}{\partial V_{\mathrm{g}_j}}$, for $i$ indexing a set of dots $\{d\}$, and $j$ indexing a set of gates $\{g\}$. Thus, the lever arm matrix can be non-square and has dimensions $n\times m$, for $n\in\{d\}$, and $m\in\{g\}$.

From our simulation results, we find that the estimated lever arm is, within the basis of left dot, right dot and$\{\mathrm{LP, RP, LB, RB, TB}\}$ gates:

\begin{equation}
    \boldsymbol{\alpha}=\left(\begin{matrix}
0.216 & 0.013 & 0.251 & 0.032 & 0.026 \\
0.013 & 0.216 & 0.032 & 0.251 & 0.026 \\
\end{matrix}\right)
\end{equation}

\noindent We considered the effects of all gates, besides the accumulation gates, to verify if potentially another gate being directly connected would have given better results. But as we have pointed out in Section \ref{sec:introduction}, this would require a differential lever arm, and thus it is clear that this would signify that the plunger gates are the optimal choice for connecting to resonator directly. These results indicate if we use the plunger gates to directly couple to the lever arm of $21.6 \%$, that we could expect a vacuum Rabi splitting of $\frac{g}{2\pi}\approx 68$ MHz.

The experimental results produced the following lever arm values:

\begin{equation}
    \boldsymbol{\alpha}=\left(\begin{matrix}
0.270 & 0.051  \\
0.049 & 0.269  \\
\end{matrix}\right)
\end{equation}

\begin{figure}
    \centering
    \includegraphics[width=1\linewidth]{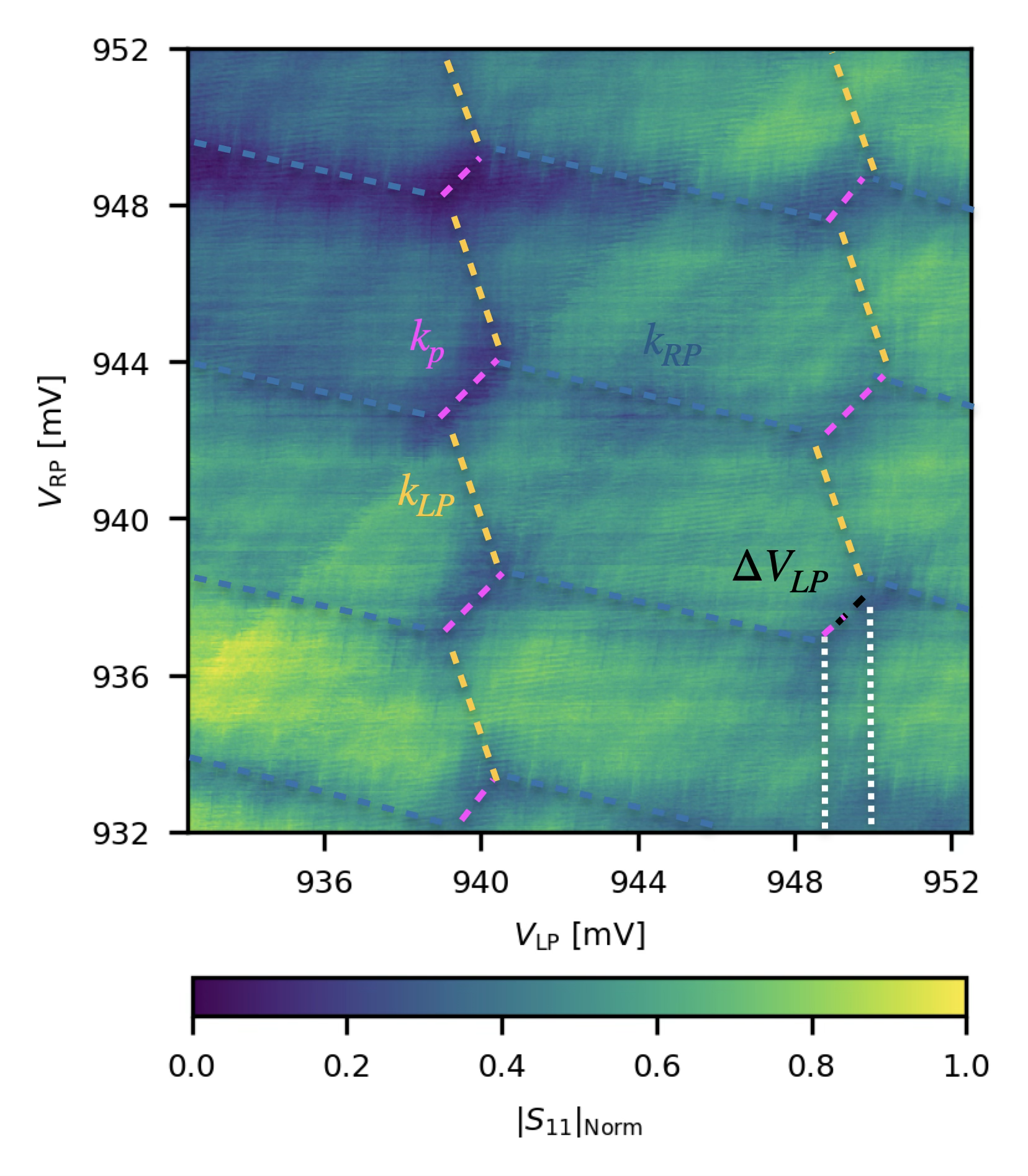}
    \caption{Experimentally obtained stability diagram when the device was tuned into a DQD. The slopes of the charging lines relate to the lever arm matrix elements as can be seen in Appendixs \ref{sup:leverarms}. We find that the plunger gates have an effect of $27\%$ and the cross capacitance is $5\%$, this was in good agreement with the simulation results. The normalized reflection coefficient is plotted.}
    \label{fig:StabDia}
\end{figure}

\noindent Experimentally we only verified the plunger gates contribution to the electrochemical potential and neglected the other gates. We estimated these lever arms by geometric considerations of the stability diagram when the device was tuned as a DQD \cite{li22,oa23,ro25}. An example of the stability diagram used to estimate these lever arms experimentally can be seen in Figure \ref{fig:StabDia}; the magnitude of the  reflection coefficient is normalized such that $|S_{11}|_\mathrm{Norm}=\frac{|S_{11}|-\mathrm{Min}(|S_{11}|)}{\mathrm{Max}(|S_{11}|)-\mathrm{Min}(|S_{11}|)}$. The details of this calculation can be seen in the Appendix \ref{sup:leverarms}. This would signify a vacuum Rabi splitting of $\frac{g}{2\pi}\approx 88$ MHz.

We note that the difference of around $33\%$ seen in the simulation to the experimental results is very likely due to the Al2O3 gate oxide. This is as the atomic layer deposition method of building the alumina is not always very well controlled or consistent, so we have seen the oxide thickness varying from $70\%-90\%$ of what we expect based on the cycles. Measurements of this gate oxide thickness indicate it is roughly $3.5$ nm, which signifies the larger experimentally observed lever arm values. Furthermore, to corroborate this difference, simulations were done with no gate oxide and produced results that had larger lever arms than what we experimentally measured. Thus, we have a strong indication that this discrepancy arises, quite likely, due to the oxide thickness and not a fault of the simulation method.

\subsection{\label{ssec:LAEng}Lever Arm Engineering}

We demonstrate how key geometric parameters affect the gate lever arm and the cross capacitive coupling by changing the SiGe cap thickness, the gate oxide thickness, and the total area of the gates and run such simulations in MaSQE. This is what we present as a method of lever arm engineering based on electrostatic simulation results. The simplest illustration of the expected change for the lever arm values comes from considering the metallic gates to form one plate of a parallel plate capacitor and the 2DEG at the QW interface to the SiGe cap as the other plate, the heterostructure and any oxide between the heterostructure of the gates represent the dielectric material of permittivity $\epsilon$ of the capacitor: $C=\frac{\epsilon A}{d}$. Under this picture, the closer the plates are to each other, smaller $d$, for a given dielectric separating them, the larger the capacitance. Similarly, the larger the area $A$, the larger the capacitance. Thus, as a simple illustration a thinner SiGe cap should present an increase in the lever arm and a thinner gate oxide will do so similarly. We list the lever arm matrix for the case of varying the SiGe cap in Table \ref{tb:sige}, where the first entry is the final design we implemented in fabrication as it was the cap thickness we had available to us. We see that a thinner SiGe cap has a dramatic increase on the lever arm of the plunger gates, we see also a possible decrease in the cross coupling. This could be considered as the fringe field effects of one gate to another are mitigated as the distance of the gate electrodes to the QW allow for less fringe fields lines to interact with with the QW not directly underneath it.

In Table \ref{tb:go}, we vary the gate oxide from the planned fabrication of 5 nm down to having no gate oxide. We see a slight increase in the plunger gate lever arm, yet not as drastic as the SiGe cap, primarily as this thickness difference is much smaller. We had planned to use this gate oxide to ensure we would not have leaky gates and with the simulations suggesting not a drastic decrease of our key parameter we wanted to optimize, we considered this a reasonable tradeoff to have high yield.

One primary consideration to get a larger gate lever arm is simply to make the gate area as large as possible while still being able to confine a discrete number of charge carriers. This is where the MaSQE simulations allowed us to have a good sense of both the confinement of our designs and if we could optimize the lever arm. Table \ref{tb:A} shows the decided gate geometry where the plunger gates have a diameter of $d_\mathrm{plunger}=100$ nm, which gave us an increase of at least twice what our group's previous devices lever arms were, and could still confine discrete single electrons within simulation. This further demonstrates against another simulation where the entire device geometry is scaled to have half the total area, such that the diameter of the plunger gate is now $d_\mathrm{plunger}=50$ nm. With just a half of the total area we see a dramatic decrease in the lever arm.

This illustrates the utility of lever arm engineering to achieve either an maximized or even minimized lever arm if so desired. This also demonstrates that cross coupling terms, though more complicated to engineer and control, can still be effected somewhat easily by material and fabrication choices.

\begin{table}[t]

\caption{Comparison of lever arm matrix as a function of the SiGe spacer cap thickness. Here the gate oxide is fixed to 5 nm for all simulations.}
\label{tb:sige}
\begin{ruledtabular}
\begin{tabular}{c c c}
 SiGe Cap $50$ nm & SiGe Cap $30$ nm & SiGe Cap $20$ nm \\
\hline
$\begin{pmatrix}
0.216 & 0.013 \\
0.013 & 0.216 
\end{pmatrix}$ &
$\begin{pmatrix}
0.346 & 0.013 \\
0.014 & 0.347
\end{pmatrix}$ &
$\begin{pmatrix}
0.437 & 0.011 \\
0.011 & 0.437
\end{pmatrix}$  \\
\end{tabular}
\end{ruledtabular}
\end{table}

\begin{table}[t]

\caption{Comparison of lever arm matrix as a function of the gate oxide thickness. Here the SiGe cap is fixed to 50 nm for all simulations.}
\label{tb:go}
\begin{ruledtabular}
\begin{tabular}{c c c}
 Al$_2$O$_3$ $5$ nm & Al$_2$O$_3$ $2.5$ nm & Al$_2$O$_3$ $0$ nm \\
\hline
$\begin{pmatrix}
0.216 & 0.013 \\
0.013 & 0.216 
\end{pmatrix}$ &
$\begin{pmatrix}
0.237 & 0.014 \\
0.014 & 0.238
\end{pmatrix}$ &
$\begin{pmatrix}
0.239 & 0.009 \\
0.009 & 0.239
\end{pmatrix}$  \\
\end{tabular}
\end{ruledtabular}
\end{table}

\begin{table}[t]

\caption{Comparison of lever arm matrix as a function of the gate width, where $d_\mathrm{plunger}$ is the diameter of the plunger gate. Here the SiGe cap is fixed to 50 nm, and the gate oxide is fixed to 5 nm for all simulations.}
\label{tb:A}
\begin{ruledtabular}
\begin{tabular}{c c}
 $d_\mathrm{plunger}=100$ nm & $d_\mathrm{plunger}=50$ nm \\
\hline
$\begin{pmatrix}
0.216 & 0.013 \\
0.013 & 0.216 
\end{pmatrix}$ &
$\begin{pmatrix}
0.069 & 0.006 \\
0.006 & 0.071
\end{pmatrix}$ \\
\end{tabular}
\end{ruledtabular}
\end{table}

\subsection{\label{ssec:SNR} Signal-to-Noise Ratio}

\begin{figure*}
    \centering
    \includegraphics{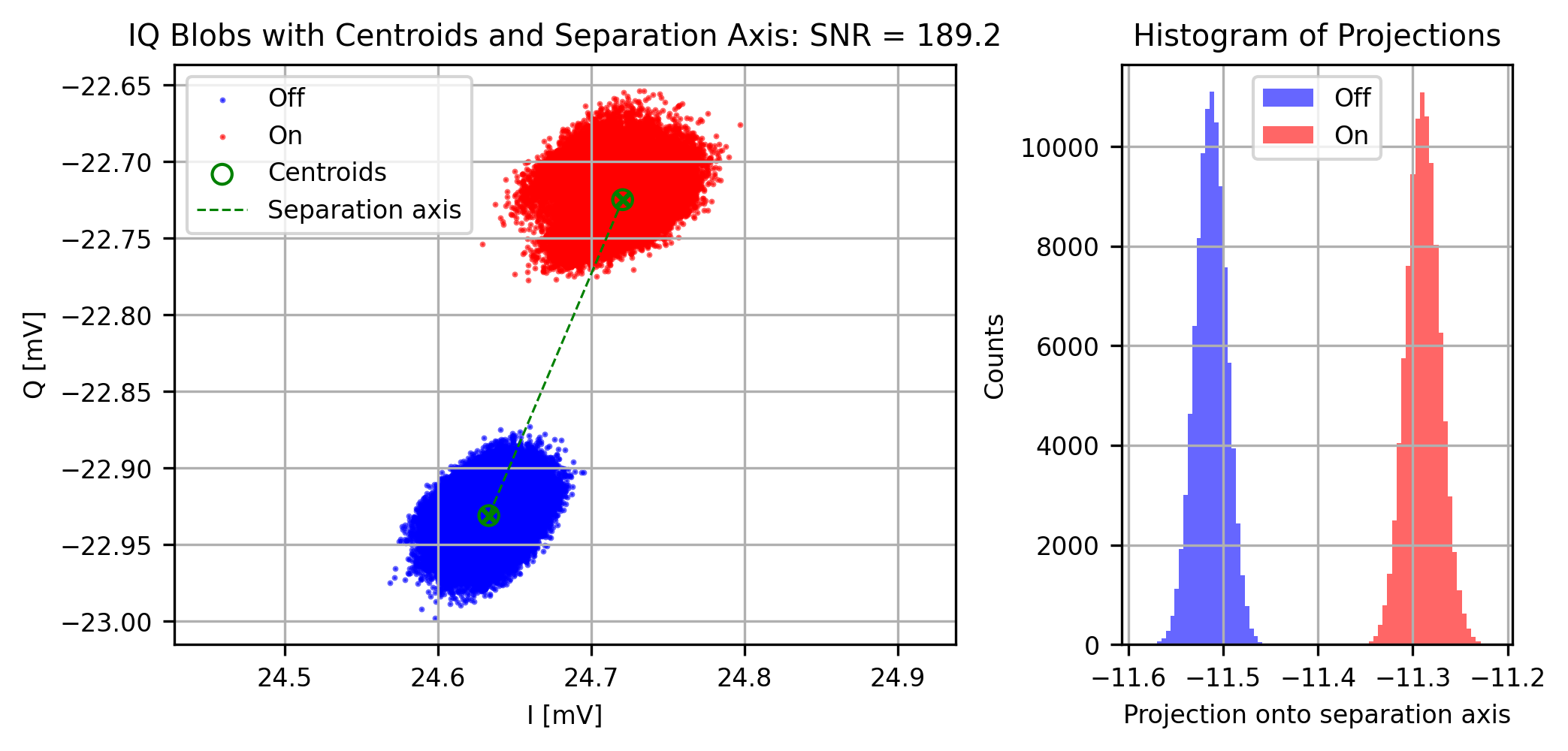}
    \caption{Left panel demonstrates IQ blobs which visually show the centroid and the separation axis to show how one can attain a geometric SNR estimate based on this. The right panel shows the projection of the distributions onto the separation axis, giving a histogram of the distributions in one-dimension. The separation axis is no longer a voltage space measure, but acts as a distance measure for all data points projected and collapsed onto the centroid joining the two distributions. This allows for a measure to be used in Equation \ref{eq:SNR}.}
    \label{fig:centroid}
\end{figure*}

We use a geometric means to quantify our systems signal to noise ratio (SNR). This is by taking a statistically large collection of data points on an interdot charge transition line, and similarly taking a large collection of data points off an interdot charge transition line in a stable point in the charge stability diagram. These two datasets gives us an on and off state that can be used to quantify the signal to the background and estimate a SNR of the system. We do this by a method referred to as IQ blobs, where the signal from the IQ-mixer is plotted in the $IQ$-plane from the on and off states. This gives us two two-dimensional distributions. How well these two distribution can be identified gives a geometric quantification of the system's SNR. We do this by collecting data for $10^5$ samples at various integration times $\tau$. An example of one set of IQ blobs can be seen in Figure \ref{fig:centroid} and the full data set can be seen in  Figures \ref{fig:IQBlobs} and \ref{fig:IQHist}.

\begin{figure*}[t]
    \centering
    \includegraphics{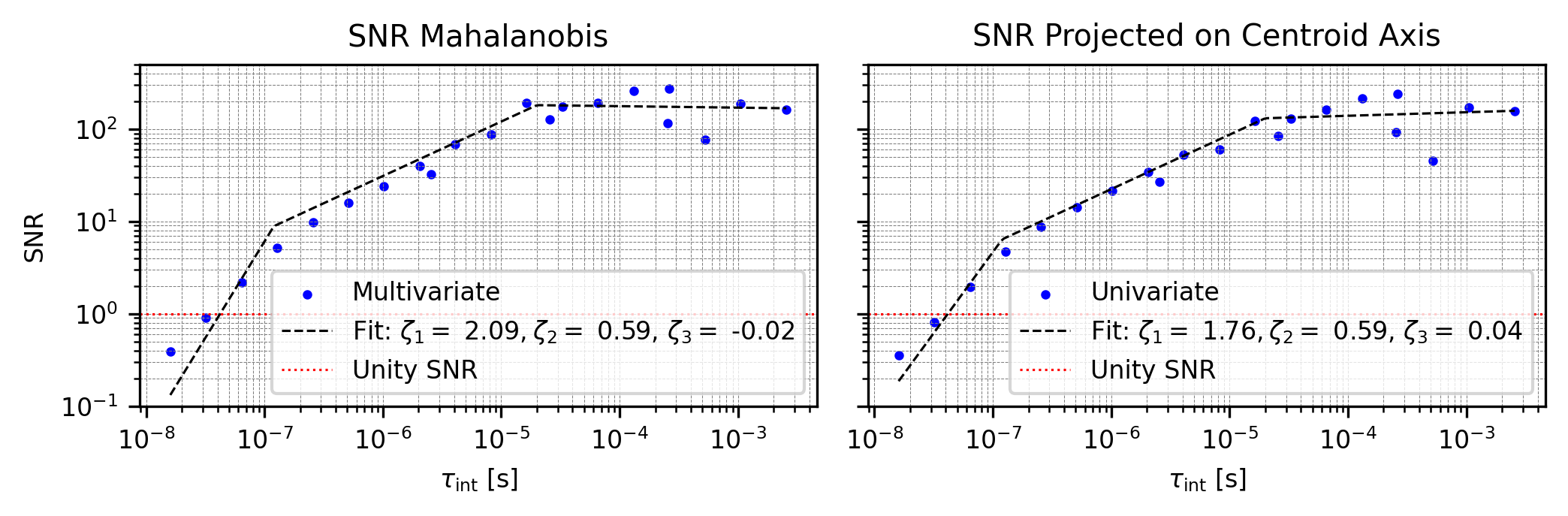}
    \caption{Signal-to-noise ratio estimates for both the Mahalanobis distance (left panel), Equation \ref{eq:MSNR}, and for the standard univariate distance (right panel), Equation \ref{eq:SNR}. We find a minimum integration time of 34.54 ns, which corresponds to a unity SNR response.}
    \label{fig:SNRest}
\end{figure*}

If one considers a univariate distribution for the on and off states, then one can quantify the SNR by:

\begin{equation}
    \text{SNR} = \frac{(I_\text{on}-I_\text{off})^2+(Q_\text{on}-Q_\text{off})^2}{\frac{1}{4}(\sigma_\text{on}+\sigma_\text{off})^2}
    \label{eq:SNR}
\end{equation}

\noindent where $I,Q_\text{on,off}$ are the inphase and quadrature signal when measured in the on and off state respectively and $\sigma_\text{on,off}$ are the standard deviations of the univariate distributions of the on and off state respectively \cite{vo24}. There is a caveat to this univariate consideration though, and this is as with longer integration times, $\tau$, the effects of low frequency charge noise come into effect and smear out the distributions. This is particularly true of the on state, which shows much less of a univariate distribution behavior for longer integration times. One can make a more sophisticated analysis by considering a multivariate distance to quantify the system's SNR. This is by considering the Mahalanobis distance:

\begin{equation}
    \text{SNR}=d_M=\sqrt{(\vec{\mu}_\text{on}-\vec{\mu}_\text{off})^{T}\Sigma^{-1}(\vec{\mu}_\text{on}-\vec{\mu}_\text{off})}
    \label{eq:MSNR}
\end{equation}

\noindent where $\vec{\mu}_\text{on,off}$ is the separation axis vector for the on and off state respectively, and $\Sigma$ is the covariance matrix averaged over the covariance matrix of the on and off states. This multivariate distance consideration improves the analysis by considering how smeared out by charge noise the distributions may become for longer integration times. We present both methods and we see a slight improvement in the system's SNR estimate using the multivariate distance. We further analyze the noise regimes present in this plot by also fitting the data to a piecewise function:

\begin{equation}
\mathrm{SNR}(\tau) =
\begin{cases}
A \left( \dfrac{\tau}{\tau_{k1}} \right)^{\zeta_1} & \tau < \tau_{k1} \\
A \left( \dfrac{\tau_{k1}}{\tau_{k1}} \right)^{\zeta_1}\left( \dfrac{\tau}{\tau_{k1}} \right)^{\zeta_2} & \tau_{k1} \le \tau < \tau_{k2} \\
A \left( \dfrac{\tau_{k2}}{\tau_{k1}} \right)^{\zeta_2}
  \left( \dfrac{\tau}{\tau_{k2}} \right)^{\zeta_3} & \tau \ge \tau_{k2}
\end{cases}
\label{eq:noisefit}
\end{equation}

\noindent where $A$ is a normalization factor, $\tau_{k1}$ is the end of the short integration time scale, $\tau_{k2}$ is the end of the intermediate integration time scale, both ensuring continuity of the piecewise fit, $\zeta_{1,2,3}$ characterize the type of noise behavior. The result can be seen in Figure \ref{fig:SNRest}.

\begin{figure}[t]
    \centering
    \includegraphics{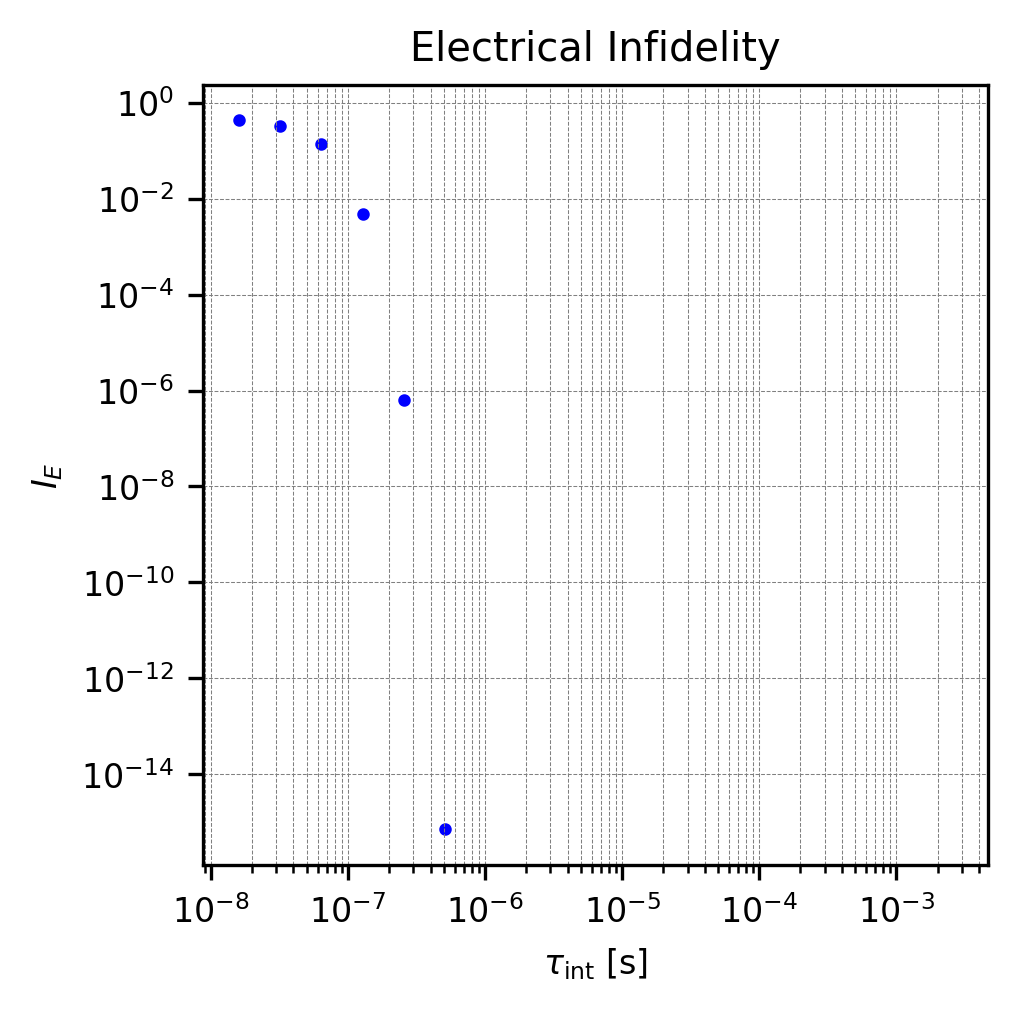}
    \caption{The electrical infidelity of the system as given by Equation \ref{eq:IE}. The results tend to zero for larger SNR values as the error function becomes close to unit value for large arguments. An electrical infidelity of $10^{-8}$ can be attained $\sim 3\times 10^{-7}$ s.}
    \label{fig:ElecInfel}
\end{figure}

We find a minimum integration time $\tau_\text{min}=34.54$  ns, which is attained by interpolation of the data acquired with the IQ blobs measurement scheme. From this we can estimate the system's charge sensitivity as $\delta q=\sqrt{\tau_\text{min}}e=1.86\times 10^{-4}\frac{e}{\sqrt{\text{Hz}}}$.

We can make a further extension of the system's SNR to an electrical infidelity as given in \cite{vo24}, which follows the form:

\begin{equation}
    I_E=1-F_E=\frac{1}{2}\left[1-\mathrm{erf}\left(\frac{\mathrm{SNR}}{2\sqrt{2}}\right)\right]
    \label{eq:IE}
\end{equation}

\noindent Where $F_E$ is the electrical fidelity and $\mathrm{erf}$ is the error function. This electrical infidelity is associated with errors of the resonator readout scheme and does not include an analysis of possible qubit errors which would need an additional consideration \cite{vo24}. The electrical infidelity can be seen in Figure \ref{fig:ElecInfel}, many of the values will plummet to zero as the error function tends to unity value for larger arguments.

The scaling of the signal-to-noise ratio (SNR) with integration time provides direct insight into the relevant noise processes of the resonator-based DQD readout. At short integration times (
$\tau_\text{int}\lesssim10^{-7}s$), the SNR increases rapidly, reflecting the dominance of uncorrelated amplifier noise consistent with white noise behavior. In the intermediate regime ($10^{-7}s\lesssim\tau_\text{int}\lesssim10^{-5}s$), the SNR follows the expected SNR $\propto \tau_\text{int}^{1/2}$ scaling, confirming that the readout is limited by Gaussian amplifier fluctuations. For longer integration times ($\tau_\text{int}\gtrsim 10^{-5}s$), the SNR saturates, indicating the onset of low-frequency noise processes such as charge noise in the double quantum dot (DQD), resonator drift, or other $1/f$-like contributions that prevent further improvement with averaging. Comparison between the Mahalanobis distance and projection onto the signal quadrature shows that axis projection mitigates correlated noise contributions, leading to a modest improvement in the asymptotic SNR. Overall, these features delineate a transition from amplifier-limited sensitivity at short times to a low-frequency noise floor at long times, consistent with the expected behavior of semiconductor charge-sensing platforms.

\subsection{\label{ssec: Charge Noise} Charge Noise}

The charge noise and charge sensitivity associated with such a system, and its comparison to the traditional QPC and SET charge sensors is of interest.
One can extract low frequency charge noise from a spectral analyzer \cite{fr16}, or by taking time traces of the resonators readout and performing the spectral density transformation on the data set \cite{be17,vo24}. We demonstrate both methods to attain a voltage power spectral density (PSD) which can be converted into a charge noise associated with our resonator measurement setup. This allows us to estimate the low frequency charge noise and attain a charge sensitivity estimate at 1 Hz from both methods. We further extend this to higher frequencies using the FPGA controller. The quantum machine's FPGA instrument allows for fast control and readout to take advantage of our system's high bandwidth.

In a similar method to how we do peak tracking for our SNR estimates, we can track the IQ-mixer signal on different points in detuning of an interdot charge transition line. If we consider the inphase signal, we may write: $\frac{\partial I}{\partial\varepsilon}=\frac{\partial I}{\partial V_g} \frac{{\partial V_g}}{{\partial \varepsilon}}=\frac{1}{\alpha_g}\frac{\partial I}{\partial V_g}$, then when this quantity is zero, there would be no tunneling events in our dot, this corresponds to a peak on an interdot charging line. When this quantity is at an extrema, then there is significant tunneling events between the left and right dot. These two points give a reference point of no tunneling events and many tunneling events respectively. Thus, if one measures the signal of both the inphase and quadrature signals at these three points, one can attain an estimate of the charge noise associated with the DQD system \cite{vi23,ba14,vo24}.

\begin{figure*}[t]
    \centering
    \includegraphics[width=\textwidth]{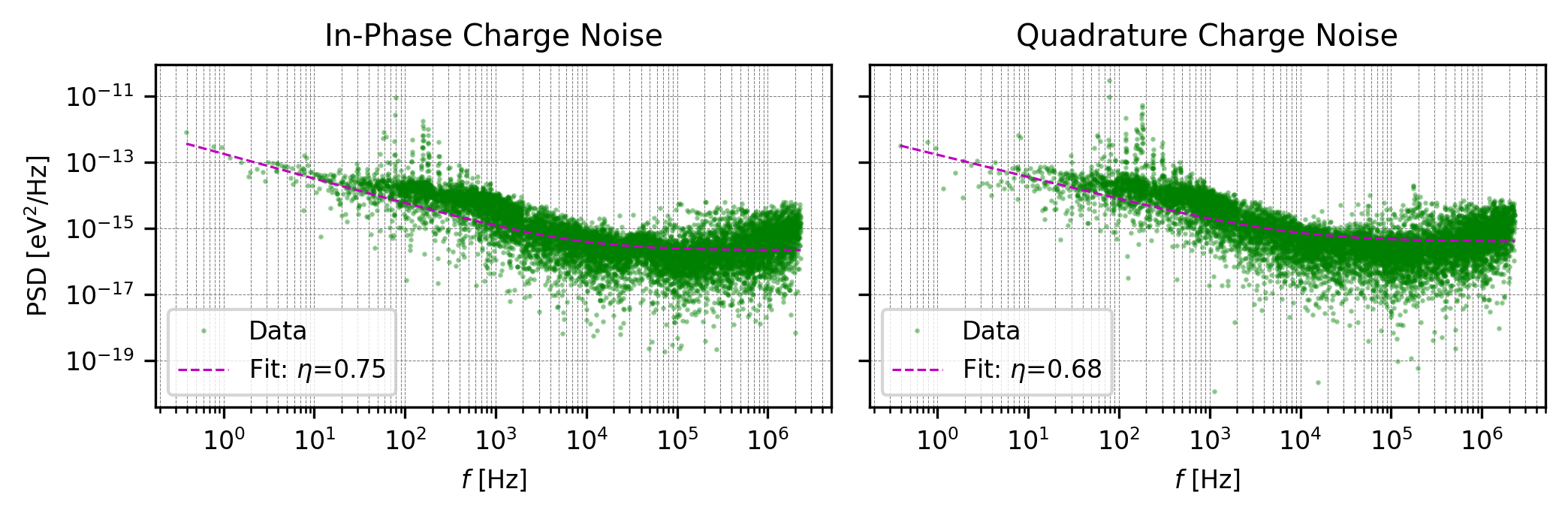}
    \caption{Charge noise estimate using Welch's method as discussed in the main text. We do this by a fast readout scheme so the estimate at 1 Hz does not present a robust estimate by itself. A longer integration time would be needed in order to properly quantify the low charge noise estimate.}
    \label{fig:chargenoise}
\end{figure*}

So we acquire such a data set, two with a considerably large amount of tunnel events in our DQD system; the extrema points. And another set with no tunneling events, where the derivative of the signal with respect to detuning is zero; a stable point in the stability diagram. Tracking the signals at these points and then converting the time domain signal into a PSD, gives a noise spectrum, $S_{II}$ and $S_{QQ}$ corresponding to the $I$ and $Q$ signals respectively. We attain a PSD of the signal by means of Welch's method \cite{we67}. The results can be seen in Figure \ref{fig:chargenoise}.

By taking the difference of the reference signal to the extrema points  signals, we attain an estimate for the charge noise associated with the DQD. The conversion from the noise spectrum PSD to the charge noise is given as:

\begin{equation}
    S_{\varepsilon\varepsilon}=\frac{S_{II}}{\left(\frac{\partial I}{\partial\varepsilon}\right)^2} \text{ and } S_{\varepsilon\varepsilon}=\frac{S_{QQ}}{\left(\frac{\partial Q}{\partial\varepsilon}\right)^2}
    \label{eq:CN}
\end{equation}

We find a $\frac{1}{f}$ behavior up to about $10$ kHz with this method. Fitting to such a $\frac{1}{f^\eta}$ model which saturates at the white noise level which is primarily limited from our amplification system. We find a slope of $0.75$ and $0.68$ for the inphase and quadrature respectively. We can further make a charge noise estimate at 1 Hz from these fits and find that: $\sqrt{S_{\varepsilon\varepsilon}(1\mathrm{ \ Hz}})= 0.42 \ \frac{\mu \text{eV}}{\sqrt{\text{Hz}}}$ and $\sqrt{S_{\varepsilon\varepsilon}(1\mathrm{ \ Hz}})= 0.41 \ \frac{\mu \text{eV}}{\sqrt{\text{Hz}}}$ from the fits for the inphase and quadrature respectively.

\section{\label{sec:discussion}Discussion}

Our results demonstrate that engineering the gate lever arm provides a robust and complementary pathway to enhancing dispersive readout in semiconductor dot–cavity systems. While prior work has emphasized impedance engineering—using high-kinetic-inductance elements or high-impedance transmission lines to boost the bare coupling $g_0$ \cite{mi17,mi18,sa18,un24}—we show that optimizing the lever arm $\alpha_g$ scales the effective coupling as $g_\text{eff}=\alpha_g g_0$. This strategy requires no exotic resonator geometries or materials and connects device design directly to measurable dispersive shifts.

We note that increasing the effective coupling $g_\text{eff}=\alpha_g g_0$ through leverarm engineering can enhance not only readout signals but also photon mediated interactions. Our approach offers a complementary pathway to impedance engineering, potentially enabling strong coupling without the need for high impedance resonators. This may support longrange spin–spin coupling via microwave photons, hybrid architectures that combine readout and entanglement channels, and integration into scalable cavity mediated networks, \cite{pe12,bo20b}. While strong coupling is not demonstrated here, we emphasize that our design strategy directly supports progress in that direction.

Maximizing $\alpha_g$ yields fast, sensitive readout: we achieve unity SNR in $\sim 34$ ns, corresponding to a $\sim14$ MHz detection bandwidth and charge sensitivity of $\sim2\times10^{-4}\ e/\sqrt{\mathrm{Hz}}$, comparable or superior to rf-QPC and rf-SET sensors \cite{fr12DI,fr12QD,vi23}. The same enhancement, however, increases susceptibility to charge noise, as it scales the PSD quadratically as seen by Equation \ref{eq:CN}, since $\frac{\partial I}{\partial\varepsilon}=\alpha_g\frac{\partial I}{\partial V_g}$. Thus, lever-arm engineering both sharpens readout and amplifies the need for material and fabrication improvements that suppress charge fluctuations.

Our results also clarify the role of bandwidth. The resonator linewidth ($\kappa/2\pi\sim 2{-}3$ MHz) sets the intrinsic cavity response, but the effective measurement bandwidth ($B_\text{eff}\sim 14.48$ MHz) reflects how quickly states can be distinguished with the amplification chain. High-fidelity error correction protocols will require consideration of both limits.

A practical advantage of our approach is that substantial lever-arm improvements can be achieved with straightforward design choices—thinning the oxide and directly overlapping resonator electrodes with plunger gates—without resorting to complex three-layer geometries. This simplicity supports scalability to larger arrays where fabrication yield and wiring density are paramount.

Future improvements include mitigating parasitic loading from unintended 2DEG accumulation to restore higher quality factors, integrating lever-arm–optimized readout with spin- and valley-based qubits to benchmark fidelities against error-correction thresholds, and combining lever-arm and impedance engineering for even stronger coupling while retaining fabrication simplicity.

In conclusion, lever-arm engineering establishes a scalable design principle for high-bandwidth dispersive readout in Si/SiGe quantum dots. By linking device geometry to coupling strength and signal strength, this work provides a clear pathway for integrating fast, high-quality measurements into semiconductor qubit architectures, advancing progress toward fault-tolerant quantum computation.

\section{\label{sec: conclusion}Conclusion}

We have realized fast, high-sensitivity dispersive readout of gate-defined Si/SiGe double quantum dots, achieving a $14.48$ MHz detection bandwidth, $\tau_\text{min}=35$ ns, and charge sensitivity of $1.9 \times 10^{-4}\ e/\sqrt{\mathrm{Hz}}$. These metrics place our device among the fastest reported charge sensors, comparable to or surpassing rf-QPC and rf-SET approaches, while relying solely on lever-arm engineering rather than high-impedance resonators. This strategy simplifies fabrication and offers a clear path to scaling in multi-qubit arrays.

Beyond readout, the same architecture enables quantitative noise spectroscopy, revealing $1/f$ charge noise up to $\sim$10 kHz and a flat spectrum at higher frequencies set by the measurement chain. This dual functionality establishes lever-arm optimization not only as a route to fast, high-fidelity measurement but also as a diagnostic tool for material and fabrication improvements.

Looking ahead, integrating this readout with spin- and valley-based qubits will enable benchmarking against quantum error correction thresholds. Combining lever-arm optimization with moderate impedance engineering could further enhance coupling while maintaining device simplicity. Taken together, these results identify lever-arm engineering as a practical and scalable design principle for high-performance readout, advancing the development of fault-tolerant quantum computation in Si/SiGe platforms.

\section*{\label{sec: acknowledgments} Acknowledgments}

We would like to acknowledge useful discussions with Jason Petta, John Dean Rooney, Hoc Ngo, and Yuwei Fan regarding device fabrication, and Mark Gyure, and Chris Anderson regarding the development of the device design and modeling and Laura Ni for assistance with analysis procedures and input on figures. Research was sponsored by the Army Research Office (ARO) and was accomplished under Grant No. W911NF-23-1-0016 (at UCLA). The views and conclusions contained in this document are those of the authors and should not be interpreted as representing the official policies, either expressed or implied, of the Army Research Office (ARO), or the U.S. Government. The U.S. Government is authorized to reproduce and distribute reprints for Government purposes notwithstanding any copyright notation herein.

\appendix

\section{\label{sec:FRO} Fast Readout of Charge Occupation}

\begin{figure*}[tbp]
    \centering
    \includegraphics{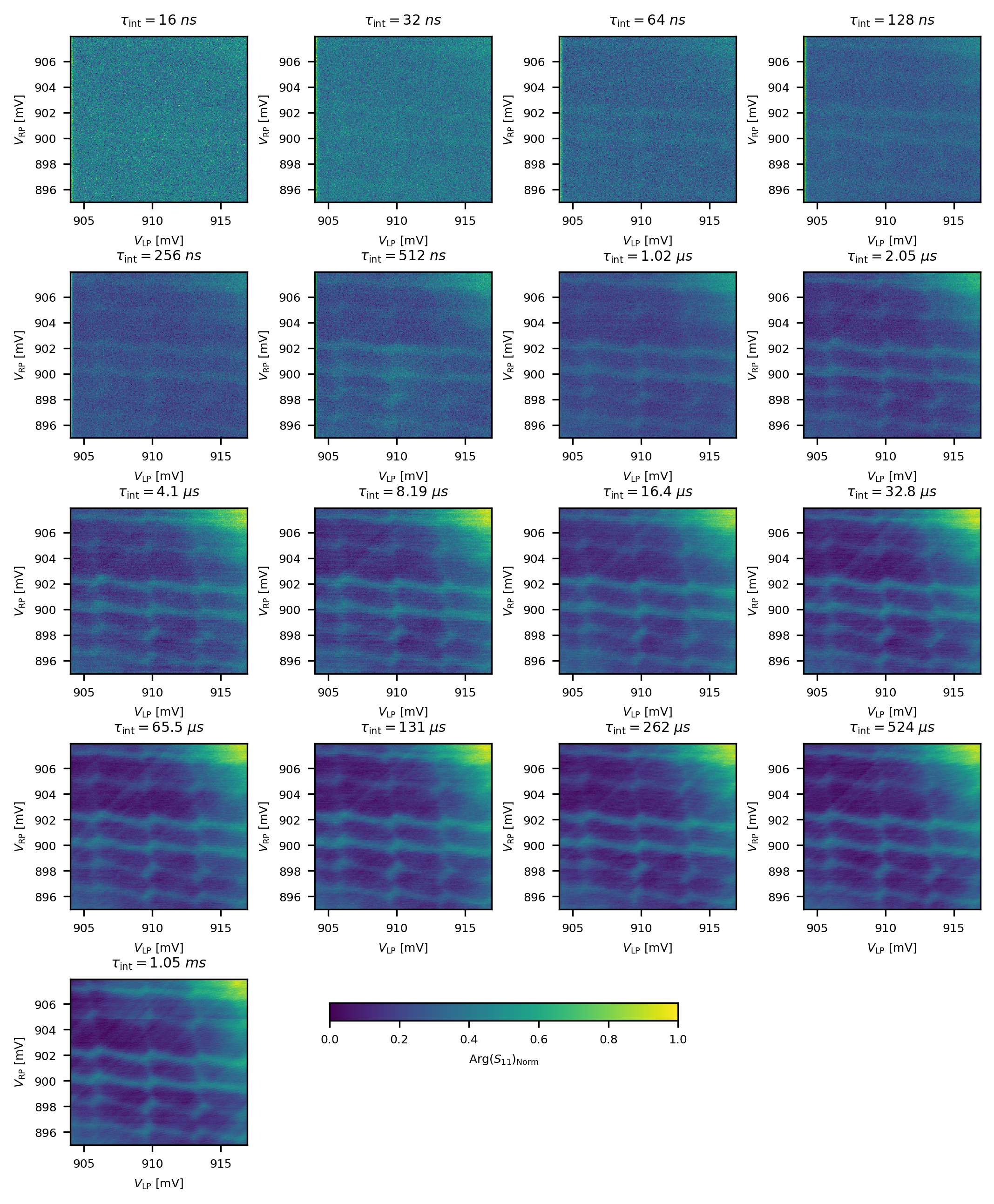}
    \caption{Stability diagrams for various integration times. Data were recorded from the resonator’s response to the changing DC bias applied to the dot’s plunger gates. Integration times varying from $2^4$ ns up to $2^{20}$ ns increasing by a factor of two each panel.} 
    \label{fig:fast}
\end{figure*}

We have been able to acquire signals showing clear transition lines with integration times on the order of 10s of ns. We can attain a high resolution stability diagram of 500 pixels by 500 pixel, with each pixel being integrated for $35$ ns in just $0.009$ s. We show example stability diagrams attained form the resonator readout with integration times varying from 16 ns,  and increasing by a factor of two each new panel, with the second panel being 32 ns, then 64 ns, 128 ns, ..., up to 1.048576 ms in  Figure \ref{fig:fast}; The argument of the reflection coefficient is plotted and is normalized in the interval $[0,1]$ such that $\mathrm{Arg}(S_{11})_{\mathrm{Norm}}=\frac{\mathrm{Arg}(S_{11})-\mathrm{Min}(\mathrm{Arg}(S_{11}))}{\mathrm{Max}(\mathrm{Arg}(S_{11}))-\mathrm{Min}(\mathrm{Arg}(S_{11}))}$ to allow for an equalized comparison between the various integration times.

\section{\label{sec:IQB} IQ Blobs Dataset Used for SNR Estimates}

\begin{figure*}[tbp]
    \centering
    \includegraphics[width=\textwidth]{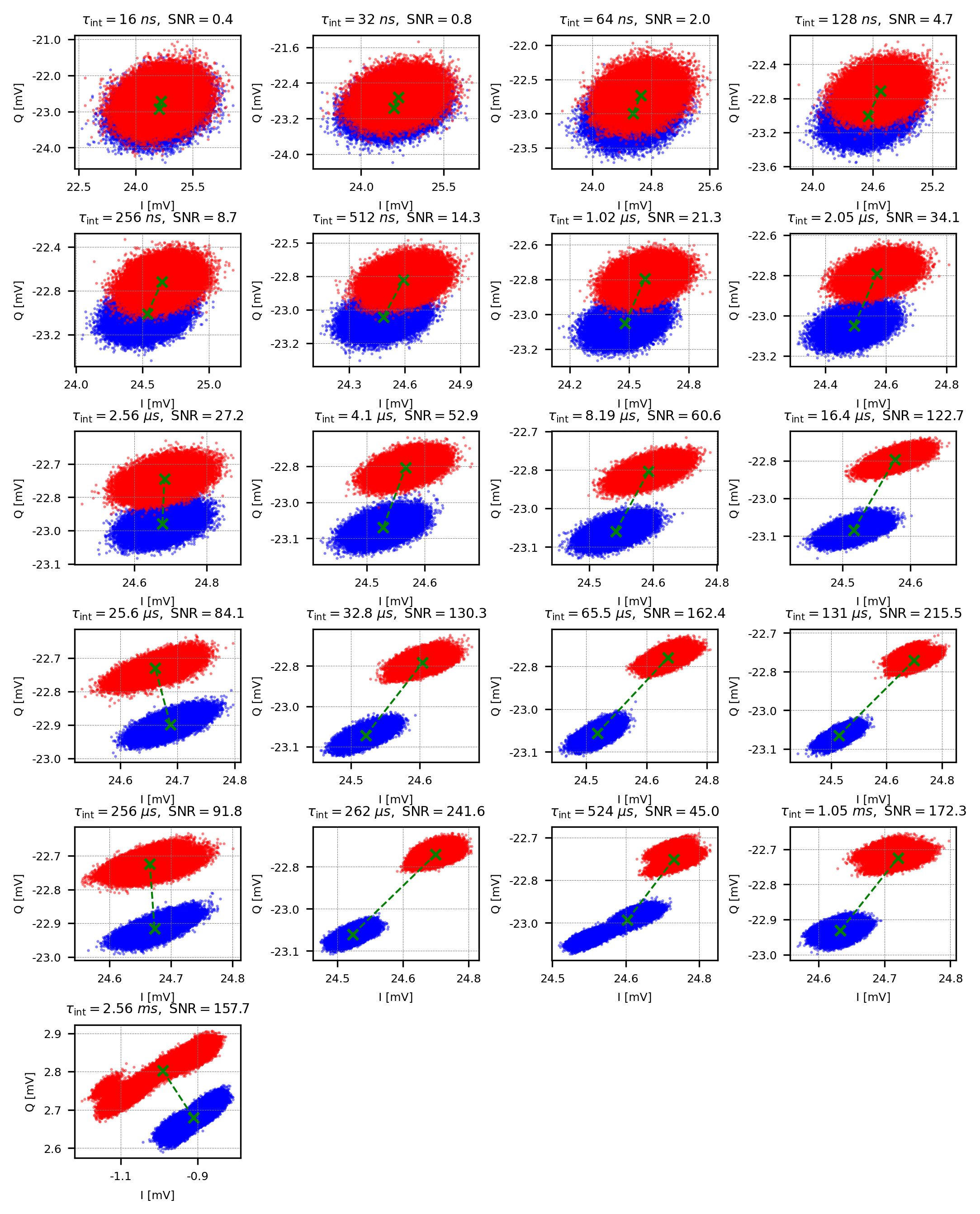}
    \caption[IQ Blobs For Various Integration Times]{The entire collection of IQ blobs used to estimate our system's SNR as seen in Figure \ref{fig:SNRest}. Times go from 16 ns up to 2.56 ms. Longer integration times suffer from low-frequency charge noise causing an elongated distribution.}
    \label{fig:IQBlobs}
\end{figure*}

\begin{figure*}[tbp]
    \centering
    \includegraphics[width=\textwidth]{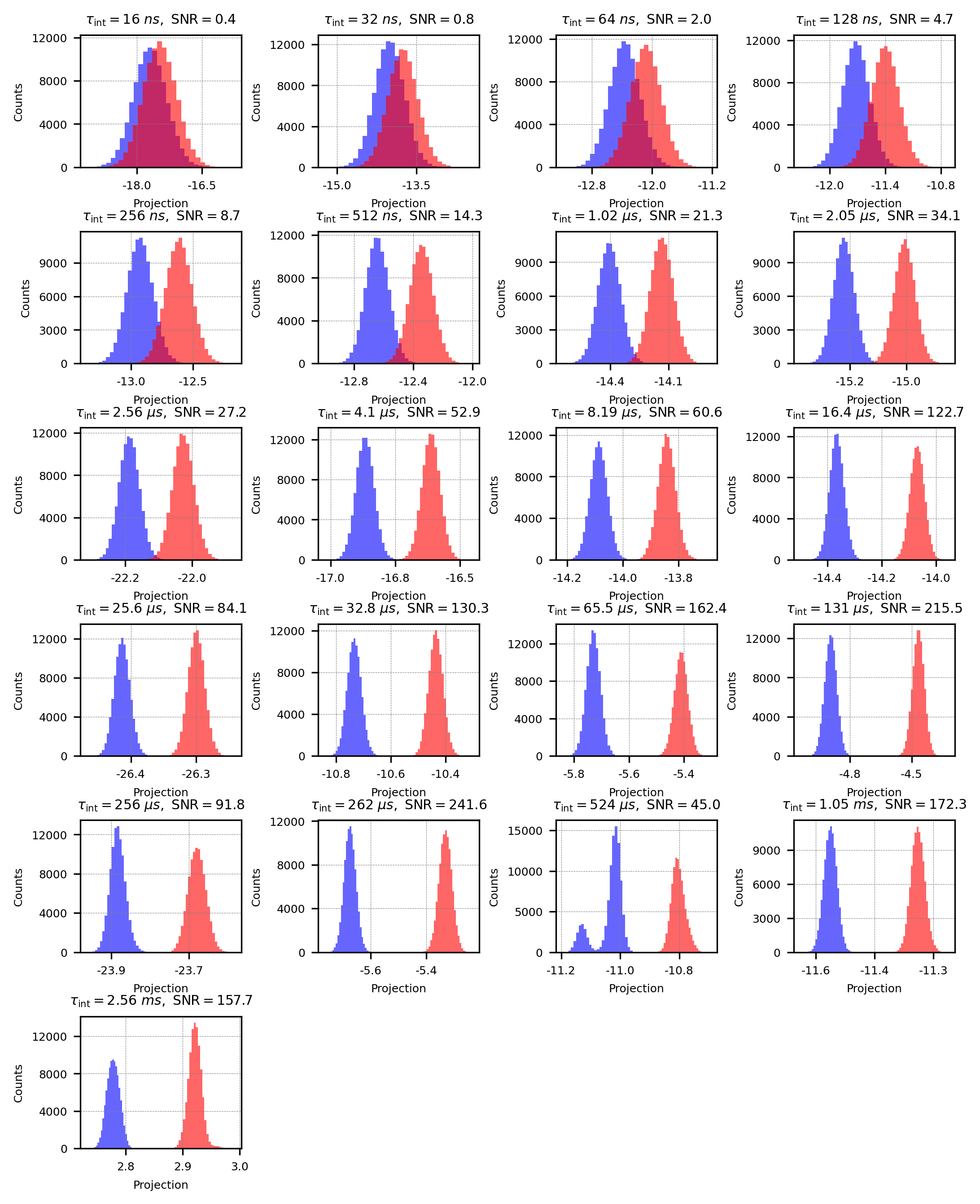}
    \caption[IQ Blobs For Various Integration Times]{The entire collection of histograms for IQ blobs projected onto the centroid axis corresponding to the set seen in Figure \ref{fig:IQBlobs}}
    \label{fig:IQHist}
\end{figure*}

The full data set used to make the estimate for the SNR as seen in the main text Figure \ref{fig:SNRest}. The integration times varied from $16$ ns up to $2.56 $ ms, the results can be seen in  Figures \ref{fig:IQBlobs} and \ref{fig:IQHist}. Low frequency charge noise can be seen to effect the distribution for longer integration times as was found also in \cite{vo24}.

%------------------Coupling Tests------------------------
\section{\label{sup:TransRes} DQD-Resonator Coupling}

\begin{figure}
    \centering
    \includegraphics[width=0.95\linewidth]{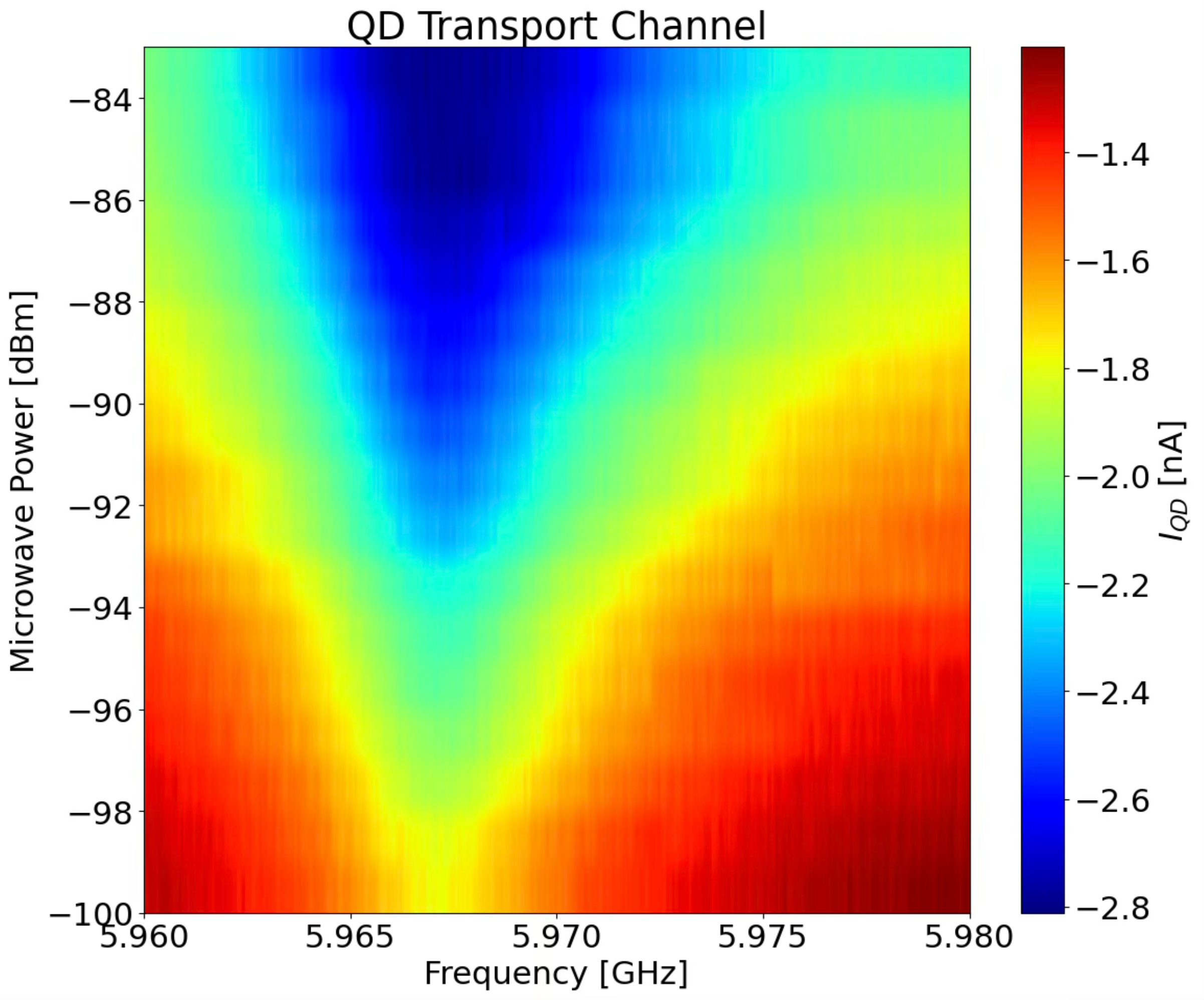}
    \caption{Increasing the applied input power to the resonant system causes a voltage excitation of the DQD channel acting as a MOSFET channel. This excitation effectively increases the gate voltage of the system causes a larger amount of current to flow. This confirms our system has a coupling between the resonator's feed-line to the DQD channel.}
    \label{fig:TransCoup}
\end{figure}

We initially test the coupling of the resonator to the DQD channel by first tuning the DQD into a regime where it acts simply as a MOSFET channel. With the gate electrodes set in such a way that an increase in the plunger gate voltages would cause a steady current to flow through the DQD. We then increase the microwave power to the resonator to see an increase in the transport current through the DQD. This becomes a voltage excitation to the gates, $V_g$, from the voltage excitation of the resonator, $V_r$, as $\delta V_r\sim \alpha_g\delta V_g \propto \sqrt{P_\text{in}}$. If one scans the drive frequency near resonance, more power goes into the system, and thus increases the voltage seen by the resonator, $\delta V_r$, becoming maximal at resonance. This, in turn, then increases the potential of the pseudo transistor gate to cause an increased flow of current through the DQD channel. This effect can be seen in Figure \ref{fig:TransCoup}. This test confirms for us that the resonator is coupled to the DQD channel and can begin testing the resonator readout.

We can also observe an effect that is primarily a heating mechanism as the resonator because energetically excited drastically by the application of an increased microwave power. When the DQD is tuned properly, we can observe a significant broadening of the coulomb peaks with increased microwave power. This can increase the electron temperature of the system and even cause the mixing chamber to start heating up when the power is too high. This is as the chip itself is dissipating the power applied to it, heating its surroundings. We must select a power which avoids this, but also gives a clear enough SNR for readout, as the SNR is proportional to the drive power, $P_\text{in}$. As stated in the main text, we operate the device around -100 dBm to -110 dBm, corresponding roughly to $10^2-10^3$ photons.

%------------------lever arms------------------------
\section{\label{sup:leverarms} Lever Arm Estimates}

The lever arm, $\alpha_\text{g}$, of a gate g relates the voltage applied to the gate, $V_\text{g}$, to the electrochemical potential, experienced in the QW by dot d, $\varepsilon_\text{d}$. It is a conversion factor from the voltage to the energy \cite{va03}. For a set of $\{n\}$ dots and a set of $\{m\}$ gates, the lever arm of the system is an $n\times m$ matrix:

\begin{equation}
\label{eq:LAmat}
\vec{\varepsilon} = -e \boldsymbol{\alpha} \vec{V}
\end{equation}

\noindent where $\vec{V}$ is the column vector of length $m$ consisting of the voltages applied to each gate $g\in \{m\}$. Considering a DQD system and just the plungers gates directly controlling the dots, we have a $2\times 2$ lever arm matrix:

\begin{equation}
\label{eq:LA2x2}
\boldsymbol{\alpha}\equiv \left(\begin{matrix}
\alpha_\text{LD,LP} & \alpha_\text{LD,RP}  \\
\alpha_\text{RD,LP} & \alpha_\text{RD,RP}  \\
\end{matrix}\right)
\end{equation}

\noindent where the labels LD(RD) indicate the left dot (right dot) and LP(RP) indicate the left plunger (right plunger) gates. The off diagonal terms give the effect of other gates onto a dot and lead to the cross capacitive effect that give rise to the honeycomb pattern seen in DQD stability diagrams \cite{li22,oa23,ro25}. This can be seen from the definition of the lever arm given in Equation \ref{eq:LA} in the main text; $\alpha_{ij}\equiv -\frac{1}{e}\frac{\partial U_{\mathrm{d}_i}}{\partial V_{\mathrm{g}_j}}$, relating the energy of the $i$-th dot to the $j$-th gate. Considering the linear system, we have:

\begin{equation}
\label{eq:LAsys}
\left(\begin{matrix}
\varepsilon_\text{LD}   \\
\varepsilon_\text{RD}  \\
\end{matrix}\right)=-e\left(\begin{matrix}
\alpha_\text{LD,LP} & \alpha_\text{LD,RP}  \\
\alpha_\text{RD,LP} & \alpha_\text{RD,RP}  \\
\end{matrix}\right)
\left(\begin{matrix}
V_\text{LD}   \\
V_\text{RD}  \\
\end{matrix}\right)
\end{equation}

When the electrochemical potentials of the dots are equal and non-zero, $\varepsilon_\text{LD}=\varepsilon_\text{RD}\ne 0$, interdot tunneling events can occur. This gives rise to a specific change in the voltages which we can state as $\left.\frac{V_{RP}}{V_{LP}}\right\rvert_{\varepsilon_{LD}= \varepsilon_{RD}}$. This change in voltage is exactly related to the slope of the interdot charging line seen in the stability diagram. Further when the electrochemical potentials of each dot is zero, dot to reservoir tunneling events can occur and for each dot. This again is a specific change in voltage for each dot which we can state as $\left.\frac{V_{RP}}{V_{LP}}\right\rvert_{\varepsilon_{LD}= 0}$ for the left dot and $\left.\frac{V_{RP}}{V_{LP}}\right\rvert_{\varepsilon_{RD}= 0}$ for the right dot.
Thus, evaluating Equation \ref{eq:LAsys} at these three points, we arrive at a set of equations which directly relate the slopes seen in a DQD charge stability diagram to the lever arm matrix elements as follows:

\begin{equation}
\begin{split}
\label{eq:Slopes}
k_{LP}&\equiv\left.\frac{V_{RP}}{V_{LP}}\right\rvert_{\varepsilon_{LD}= 0}=-\frac{\alpha_{LD,LP}}{\alpha_{LD,RP}}, \\ 
k_{RP}&\equiv\left.\frac{V_{RP}}{V_{LP}}\right\rvert_{\varepsilon_{RD}= 0}=-\frac{\alpha_{RD,LP}}{\alpha_{RD,RP}}, \\ 
k_p&\equiv\left.\frac{V_{RP}}{V_{LP}}\right\rvert_{\varepsilon_{LD}= \varepsilon_{RD}}=\frac{\alpha_{LD,LP}-\alpha_{RD,LP}}{\alpha_{RD,RP}-\alpha_{LD,RP}}
\end{split}
\end{equation}

Furthermore, assuming a finite bias of $V_{SD}$ applied to the left dot, when the electrochemical potentials of the two dots are equal, we find that:

\begin{equation}
\label{eq:LASD}
\frac{|V_{SD}|}{\Delta V_{LP}}=\alpha_{LD,LP}+k_p\alpha_{LD,RP}
\end{equation}

These four Equations, \ref{eq:LAsys} and \ref{eq:LASD}, allow us to attain each lever arm matrix element.

\bibliography{manuscript.bbl} 

\end{document}